\documentclass[3p]{elsarticle} 

\usepackage{setspace}

\usepackage{lineno,hyperref}
\modulolinenumbers[5]

\journal{International Journal of Fatigue}

\usepackage[utf8]{inputenc} 
\usepackage{graphicx,psfrag,color}
\usepackage{subfig} 
\usepackage{caption} 
\usepackage{tabularx}
\usepackage{booktabs} 
\usepackage{multirow}
\usepackage{float}

\usepackage{amsmath} 
\usepackage{amssymb} 
\usepackage{amsfonts}
\usepackage{amsxtra}
\usepackage{tensor}
\usepackage{empheq}
\usepackage{stmaryrd}
\usepackage{icomma}
\usepackage{relsize}
\usepackage{trfsigns}
\usepackage[locale=US, unit-mode=text]{siunitx}

\usepackage{xcolor}

\newcommand{\M}[1]{\underline{\underline{\mathbf #1}}} 
\newcommand{\V}[1]{\underline{\mathbf #1}} 
\newcommand*\diff{\mathop{}\!\mathrm{d}}

\usepackage{tikz}
\usetikzlibrary{fadings,shapes.arrows,shadows}
\usepackage{xparse}
\usepackage{xcolor}
\tikzfading[name=arrowfading, top color=transparent!0, bottom color=transparent!95]
\tikzset{arrowfill/.style={#1,general shadow={fill=black, path fading=arrowfading}}}
\tikzset{arrowstyle/.style n args={3}{draw=#2,arrowfill={#3}, single arrow,minimum height=#1, single arrow,
		single arrow head extend=.2cm,}}
\NewDocumentCommand{\tikzfancyarrow}{O{2cm} O{FireBrick} O{top color=OrangeRed!20, bottom color=Red} m}{
	\tikz[baseline=-0.5ex]\node [arrowstyle={#1}{#2}{#3}] {#4};
}

\usepackage{pgfplots}

\usepackage{array,booktabs,ragged2e}
\newcolumntype{R}[1]{>{\RaggedLeft\arraybackslash}p{#1}}
\newcolumntype{P}[1]{>{\centering\arraybackslash}p{#1}}

\usetikzlibrary{positioning}
\usetikzlibrary{backgrounds}
\tikzstyle{background rectangle}=[draw]

\DeclareUnicodeCharacter{2212}{--} 

\usepackage{array} 

\let\oldalign\align
\let\oldendalign\endalign

\renewenvironment{align}
{\linenomathNonumbers\oldalign}
{\oldendalign\endlinenomath}	

\begin{document}

\begin{frontmatter}

\title{Fatigue monitoring and maneuver identification for vehicle fleets using a virtual sensing approach}

\author[IFKM]{L. Heindel}
\author[IFKM,DCFR]{P. Hantschke}
\author[IFKM,DCFR]{M. Kästner\corref{cor1}}

\address[IFKM]{Technische Universität Dresden, Institute of Solid Mechanics, 01062 Dresden, Germany}
\address[DCFR]{Dresden Center for Fatigue and Reliability (DCFR), 01062 Dresden, Germany}

\cortext[cor1]{Corresponding author. E-mail: Markus.Kaestner@tu-dresden.de ; Tel.: +49 351 463-43065 ; fax: +49 351 463-37061.}

\begin{abstract}
	Extensive monitoring comes at a prohibitive cost, limiting Predictive Maintenance strategies for vehicle fleets. This paper presents a measurement-based virtual sensing technique where local strain gauges are only required for few reference vehicles, while the remaining fleet relies exclusively on accelerometers. The scattering transform is used to perform feature extraction, while principal component analysis provides a reduced, low dimensional data representation. This enables direct fatigue damage regression, parameterized from unlabeled usage data. Identification measurements allow for a physical interpretation of the reduced representation. The approach is demonstrated using experimental data from a sensor equipped eBike, which is made publicly available.
\end{abstract}

\begin{keyword}
predictive maintenance \sep fatigue monitoring \sep maneuver identification \sep scattering transform \sep soft sensing
\end{keyword}

\end{frontmatter}


\section{Introduction}

Engineering vehicles against fatigue failure is a complex problem, which consists in a comparison between the fatigue strength of the vehicle components and the expected operational loads during the lifetime of the vehicle. Many load assumptions have to be made, in order to generate a representative load collective, which reflects the expected regular usage scenarios and simultaneously takes special events into account. A detailed summary of both established procedures, as well as novel approaches in this field, is given in \textsc{Burger} et al. \cite{Burger2021}. Once a design is finished, the probability of failure for an individual component, with respect to the chosen load collective, is known. However, the  particular load history, which any individual vehicle experiences during its service life, is generally not known. This is a significant challenge for the estimation of its remaining useful life time.\\
The aim of predictive maintenance is to identify accumulating damage in critical components preemptively, i.e. before failure occurs, in order to schedule replacements by demand. This allows to better exploit real load bearing capabilities and to prevent system failures. Numerous existing works in the field of structural health monitoring (SHM) address this problem by monitoring system responses with the aim of detecting changes in the system behavior and thus inferring component damage. A detailed review of classical approaches in SHM for civil infrastructure is given by \textsc{Brownjohn} \cite{Brownjohn2007}, while a review of recent approaches involving machine learning algorithms is provided in \cite{Flah2021}. The approach of detecting changes in the system behavior is feasible, since buildings are generally designed in a redundant manner, where the failure of an individual substructure does not cause an immediate collapse of the entire structure. Motor vehicles, however, are generally build with lightweight design in mind. While a variety of crash scenarios are taken into account for reasons of passenger safety, the failure of individual components can immediately impair the roadworthiness of the vehicle. Damage due to material fatigue accumulates over a long period of time and can occur very abruptly when a critical damage threshold is reached. As a result, detecting changes in the system behavior can be too reactive for many applications.\\
A predictive maintenance approach for these applications should therefore monitor the load history at critical components in a system, in order to provide information about the fatigue damage accumulation. To conduct a fatigue damage calculation following established procedures \cite{Haibach2002}, local strain or stress information is required. The installation of a sufficient number of sensors, e.g. strain gauges, for such an elaborate monitoring is, however, uneconomic and impractical in many applications and therefore usually limited to prototypes in the design process or designated measuring vehicles, as described in \textsc{Stellmach} et al. \cite{Stellmach2021}. This is unfortunate, since service and repair of consumer vehicles could benefit immensely from such monitoring.\\
Frequency based fatigue life estimation methods \cite{Quigley2016} are commonly used in the automotive industry, since they are suited for stationary Gaussian loadings, resulting from a combustion engine and random underground excitation. A promising approach is provided by \textsc{Ugras} et al. \cite{Ugras2019}, where the fatigue life of truck components is monitored through acceleration measurements, using frequency response functions obtained by finite element simulations and frequency domain methods for fatigue estimation, implemented as an on board solution. Other possible solutions to this problem are given by virtual sensing (VS) or soft sensing approaches, which approximate unmeasured system responses or related quantities of interest based on available measurement data. This enables system monitoring through reduced sensor configurations and provides solutions in situations where quantities of interest are difficult to measure or sensor locations are inconvenient to access. Virtual sensing approaches can be fundamentally divided into model-based and data-based approaches.\\
Model-based strategies are characterized by the presence of a physics based model, which describes the relationship between the system response of interest and the measurable quantities. Examples from this category are commonly methods based on finite element simulations and experimental modal analysis. \textsc{Hjelm} et al. \cite{Hjelm2005} shows that stress histories can be obtained indirectly from strain measurements using a physical VS model, which is shown using experimental measurements from a lattice tower. Similar approaches, which also employ the combination of finite element simulation and modal analysis, are provided by \textsc{Tarpo} et al. \cite{Tarpo2020} and \textsc{Flores Terrazas} et al. \cite{FloresTerrazas2022}, where VS is used to enable fatigue monitoring of an offshore structure and wind turbine tower, respectively. In the field of motor vehicles, physical modeling approaches like the finite element method can be costly to parameterize and obtaining accurate predictions for highly dynamic and non-linear systems is generally very resource intensive.\\
In contrast, data-based approaches rely on existing datasets, which contain the relationship between measurements and responses implicitly. A variety of machine learning approaches exist in this field. In \textsc{Rouss} et al. \cite{Rouss2009}, an artificial neural network is employed to predict the behavior of a non-linear dynamic system, which is demonstrated using measurements from a vibrating platform. \textsc{Kullaa} \cite{Kullaa2015, Kullaa2019} presents various methods for both model- and data-based VS, e.g. estimating redundant sensors in a network, reducing measurement noise and providing full field information from a limited set of sensors. Data driven VS is also employed for audio transmission in \textsc{Wang} et al. \cite{Wang2020}, where noise signals at the passenger ear side are reconstructed from audio measurement at different locations in the interior. During the parameterization phase, these methods require measurements of the system response of interest. Once the system is deployed, these measurements are no longer required and instead replaced by data-based predictive models.\\ 
The aim of this paper is to provide an economic soft sensing approach for fatigue damage monitoring of vehicle fleets, which can be parameterized using only measurement data. It is based on a low dimensional data representation, which is obtained by a combination of feature extraction and dimensionality reduction. Here, the scattering transform provides a powerful tool for the extraction of fatigue related information, leading to two separate applications, namely fatigue damage regression and maneuver identification. A sensor-equipped eBike serves as a demonstrator for the presented approach, with which an experimental dataset was acquired. This dataset is made publicly available for future research.\\
The paper is structured as follows: First, necessary fundamentals of the presented approach, i.e. Scattering Transformation and PCA are presented. On this basis, the low-dimensional data representation is elaborated. Subsequently, the experimental setup is presented and the resulting data sets are described in detail. As the first application, an approach for direct damage monitoring using a reduced sensor configuration is presented. Secondly, the data representation is used in a discrete maneuver identification example. Afterwards, the most essential results of this study are evaluated and discussed. The paper concludes with a short summary and an outlook on possible extensions and improvements of the presented methods.\\

\section{Material and Methods} \label{sec:model_setup}
In this paper, a combination of the scattering transform and principal component analysis (PCA) is used to extract and compress useful sensor information. In the following explanations, vectors $\V v$ and matrices $\M M$ are denoted by single and double underscores, respectively.

\subsection{Fourier and Wavelet transform}
Given a time series $x(t)$, the Fourier transform of this time series
\begin{align}
	\hat{x}(\omega) = \int_{-\infty}^{\infty} x(t) e^{-i2\pi \omega t} \diff t
\end{align}
describes the frequency content of $x(t)$, but does not include information about the specific point in time, at which an observed frequency occurred. Since many applications exist, where information from both time and frequency domains plays an important role, additional analysis tools where developed. The Short Time Fourier transform (STFT) computes the Fourier transform of signals over small time windows, leading to a spectrogram which displays frequency and time information simultaneously with homogeneous resolution. In a similar manner, the continuous wavelet transform (CWT) is obtained by computing the correlation
\begin{align}
	x_\psi(a,b) = \frac{1}{|a|^\frac{1}{2}} \int_{-\infty}^{\infty} x(t) \overline{\psi\left(\frac{t-b}{a}\right)} \diff t = \frac{1}{|a|^\frac{1}{2}} x(t) \star \overline{\psi_{a,b} (t)} 
\end{align}
between $x(t)$ and a wavelet basis
\begin{align}
	\psi_{a,b} (t) = \psi\left(\frac{t-b}{a}\right), \label{eq:wavelet_basis}
\end{align}
which is constructed from a mother wavelet $\psi (t)$ using a dilation factor $a$ and a shifting parameter $b$. The operator $\star$ refers to the correlation operator
\begin{align}
	f(\tau) \star g(\tau) = \int_{-\infty}^{\infty} f(\tau) \overline{g(t+\tau)} \diff t.
\end{align}
The CWT results in a scalogram, which is similar to the spectrogram in the sense that it contrains information in the frequency and time domain. The main difference is that the scalogram leads to an adaptive resolution, where low frequencies in $x(t)$ are resolved with a high resolution in the frequency domain, but a low resolution in time, and high frequencies in $x(t)$ feature a high resolution in the time domain and a low resolution in frequency.\\

\subsection{Scattering transform} \label{sec:Sct}
The Scattering transform, introduced in \cite{Bruna2013,Anden2014,Mallat2016}, extracts signal invariants from a time series in order to provide useful features to discriminate between different classes of signals, i.e., a classification task. It has been employed in different temporal and spatial classification problems, where similar results to state of the art deep neural network approaches were achieved without necessitating a data intensive training process. In \textsc{Seydoux} et al. \cite{Seydoux2020}, it provides the basis for an unsupervised clustering framework which detects seismic signals. The transform is explained in great detail in \cite{Anden2014} and only briefly reviewed here.\\
\begin{figure}
	\begin{center}
%
%
%
%

\begin{tikzpicture}


\def\Scale{1.35};
\def\Xorigin{0};
\def\Yorigin{0};
\def\TimeseriesLength{6*\Scale};
\def\TimeseriesWidth{0.618*0.7*\Scale};
\def\SegmentStart{1*\Scale};
\def\SegmentLength{1*\Scale};

\def\ScatteringXorigin{\Xorigin+\SegmentStart+0.5*\SegmentLength};
\def\ScatteringYorigin{\Yorigin};
\def\LevelOneHight{1*\Scale};
\def\LevelOneDistance{2*\Scale};

\def\LevelTwoHight{1*\Scale};
\def\LevelTwoDistance{0.5*\Scale};

\def\XDescription{\ScatteringXorigin+1.75*\LevelOneDistance};
\def\XDescriptionDistance{0.5*\Scale};
\def\DescriptionYLevelZero{\Yorigin+0.5*\TimeseriesWidth-0.6*1.2*\LevelOneHight};
\def\DescriptionYLevelOne{\ScatteringYorigin-\LevelOneHight-0.5*\LevelOneHight};
\def\DescriptionYLevelTwo{\ScatteringYorigin-\LevelOneHight-\LevelTwoHight-0.5*\LevelTwoHight};
\def\DescriptionMinimumWidth{3.7cm};
\def\DescriptionTextWidth{3.5cm};

\def\XCoefficientOrigin{\XDescription+9*\XDescriptionDistance};
\def\YCoefficientOrigin{\DescriptionYLevelZero+0.5*\LevelOneHight};
\def\CoefficientLength{0.3*\Scale};
\def\CoefficientWidth{0.3*\Scale};
\def\CoefficientZeroRatio{1};
\def\CoefficientOneRatio{3};
\def\CoefficientTwoRatio{5};
\def\YCoefficientZeroNode{\YCoefficientOrigin -0.5*\CoefficientZeroRatio*\CoefficientWidth};
\def\YCoefficientOneNode{ \YCoefficientOrigin -1.0*\CoefficientZeroRatio*\CoefficientWidth -0.5*\CoefficientOneRatio*\CoefficientWidth};
\def\YCoefficientTwoNode{ \YCoefficientOrigin -1.0*\CoefficientZeroRatio*\CoefficientWidth -1.0*\CoefficientOneRatio*\CoefficientWidth -0.5*\CoefficientTwoRatio*\CoefficientWidth};

\node at (\ScatteringXorigin, \ScatteringYorigin+1.5*\TimeseriesWidth) {$x(t)$};
\node at (\ScatteringXorigin-0.85*\LevelOneDistance, \DescriptionYLevelZero+0.1*\LevelOneHight) {$|(\cdot) \star \lambda_1|$};
\node at (\ScatteringXorigin-1*\LevelOneDistance-2*\LevelTwoDistance, \DescriptionYLevelOne) {$|(\cdot) \star \lambda_2|$};

\node[rectangle,draw=gray!100,fill=gray!0] (D0) at (\XDescription+\XDescriptionDistance, \DescriptionYLevelZero) {$(\cdot) \star \phi$};
\node[rectangle,draw=gray!100,fill=gray!0] (D1) at (\XDescription+\XDescriptionDistance, \DescriptionYLevelOne) {$(\cdot) \star \phi$};
\node[rectangle,draw=gray!100,fill=gray!0] (D2) at (\XDescription+\XDescriptionDistance, \DescriptionYLevelTwo) {$(\cdot) \star \phi$};

\node[anchor=west,minimum width=0.45*\DescriptionMinimumWidth,text width=0.45*\DescriptionTextWidth] (S0) at (\XDescription+\XDescriptionDistance+1.5*\XDescriptionDistance, \DescriptionYLevelZero) {$S_0  =   x \star \phi$};
\node[anchor=west,minimum width=0.7*\DescriptionMinimumWidth,text width=0.7*\DescriptionTextWidth] (S1) at (\XDescription+\XDescriptionDistance+1.5*\XDescriptionDistance, \DescriptionYLevelOne) {$S_1^i =  |x \star \lambda_1^i| \star \phi$};
\node[anchor=west,minimum width=\DescriptionMinimumWidth,text width=\DescriptionTextWidth] (S2) at (\XDescription+\XDescriptionDistance+1.5*\XDescriptionDistance, \DescriptionYLevelTwo) {$S_2^{ij} = ||x \star \lambda_1^i| \star \lambda_2^j|\star \phi$};
\draw[->, color=gray, thin] (D0) -- (S0);
\draw[->, color=gray, thin] (D1) -- (S1);
\draw[->, color=gray, thin] (D2) -- (S2);

\draw[draw=black, thick] (\Xorigin+\SegmentStart,\Yorigin) rectangle ++(\SegmentLength,\TimeseriesWidth);

\draw[color=gray, thin] (\Xorigin+\SegmentStart+0.5*\SegmentLength, \Yorigin+0.5*\TimeseriesWidth) -- (\Xorigin+\SegmentStart+0.5*\SegmentLength+1.2*1.2*\LevelOneHight, \DescriptionYLevelZero) -- (D0);

\foreach \x in {-1,...,1}
{
	\node[circle, inner sep=4pt*\Scale, draw=black] (\x 1) at (\ScatteringXorigin+\x*\LevelOneDistance, \ScatteringYorigin-\LevelOneHight) {};
	\draw[color=gray, thin] (\ScatteringXorigin+\x*\LevelOneDistance, \ScatteringYorigin-\LevelOneHight) -- (\ScatteringXorigin+\x*\LevelOneDistance+0.8*\LevelOneHight, \DescriptionYLevelOne) -- (D1);
	\draw[->] (\ScatteringXorigin, \ScatteringYorigin) -- (\x 1);
	
	\foreach \xx in {-1,...,1}
	{
		\node[circle, inner sep=4pt*\Scale, draw=black] (\xx 2) at (\ScatteringXorigin+\x*\LevelOneDistance+\xx*\LevelTwoDistance, \ScatteringYorigin-\LevelOneHight-\LevelTwoHight) {};
		\draw[color=gray, thin] (\ScatteringXorigin+\x*\LevelOneDistance+\xx*\LevelTwoDistance, \ScatteringYorigin-\LevelOneHight-\LevelTwoHight) -- (\ScatteringXorigin+\x*\LevelOneDistance+\xx*\LevelTwoDistance+0.8*\LevelTwoHight, \DescriptionYLevelTwo) -- (D2);
		\draw[->] (\x 1) -- (\xx 2);
	}
}

\node at (\XCoefficientOrigin + 0.5*\CoefficientLength, \YCoefficientOrigin+0.8*\CoefficientWidth) {$\V S$};
\draw[draw=black] (\XCoefficientOrigin, \YCoefficientOrigin) rectangle ++(\CoefficientLength,-\CoefficientWidth*\CoefficientZeroRatio);
\draw[draw=black] (\XCoefficientOrigin, \YCoefficientOrigin-\CoefficientWidth*\CoefficientZeroRatio) rectangle ++(\CoefficientLength,-\CoefficientLength*\CoefficientOneRatio);
\draw[draw=black] (\XCoefficientOrigin, \YCoefficientOrigin-\CoefficientWidth*\CoefficientZeroRatio-\CoefficientWidth*\CoefficientOneRatio) rectangle ++(\CoefficientLength, -\CoefficientLength*\CoefficientTwoRatio);

\draw[->, color=gray, thin] (S0.east) -- (\XCoefficientOrigin+0.5*\CoefficientLength,\YCoefficientZeroNode);
\draw[->, color=gray, thin] (S1.east) -- (\XCoefficientOrigin+0.5*\CoefficientLength,\YCoefficientOneNode);
\draw[->, color=gray, thin] (S2.east) -- (\XCoefficientOrigin+0.5*\CoefficientLength,\YCoefficientTwoNode);

\end{tikzpicture}

		\caption[scattering]{The scattering transform applies multiple layers of wavelet convolutions $(\cdot)\star \lambda_w$ and modulus operations $|(\cdot)|$ to an input signal $x(t)$. The vector of scattering coefficients $\V S$ is structured in sections corresponding to each layer. These sections contain an entry per dilated wavelet, time averaged via a low pass filter $\phi$.}
		\label{fig:scattering}
	\end{center}
\end{figure}
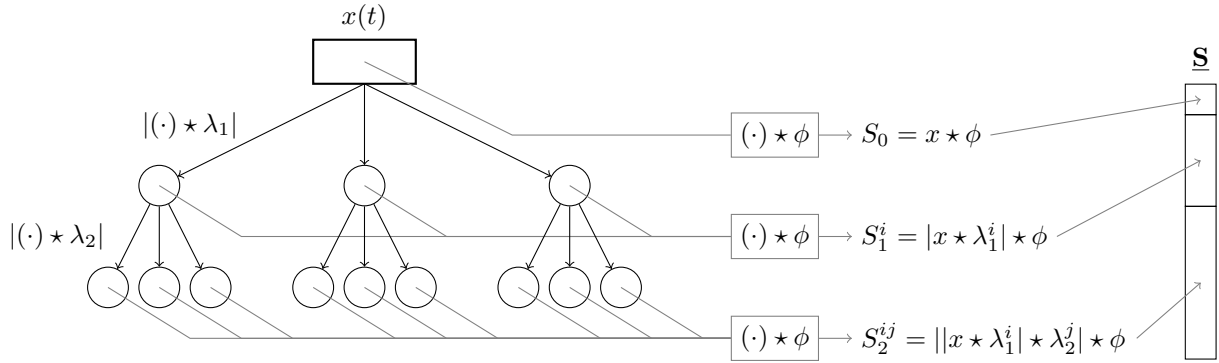
The architecture of the scattering transform is depicted in \autoref{fig:scattering}. It bears many similarities to convolutional neural networks and is designed to extract features from $x(t)$, which are both translation invariant and stable to time warping deformation. This is achieved by iteratively computing multiple layers of wavelet convolutions $(\cdot)\star \lambda_w$ and modulus operations $|(\cdot)|$, where the index $w$ denotes the layer index.\\
The wavelet basis used in the first scattering layer $\lambda_1$ is similar to the basis $\psi_{a,b} (t)$ introduced in \autoref{eq:wavelet_basis} in the sense that a mother wavelet $\psi(t)$ is dilated
\begin{align}
	\psi_{\lambda_1}(t) = \lambda_1 \psi(\lambda_1 t),
\end{align}
by a set of dilation factors
\begin{align}
	\lambda_1 \in \{ 2^{k/Q}|k \in \mathbb{Z}\} ,
\end{align}
defined by a given number of wavelets per octave $Q$. Afterwards, a modulus operation is applied and the first layer of the transform $|x(t)\star\lambda_1|$ is complete. The information contained in this layer is similar to the information in a spectrogram or scalogram.\\
For the next layer, the process is repeated iteratively using the wavelet basis
\begin{align}
	\psi_{\lambda_2}(t) = \lambda_2 \psi(\lambda_2 t),
\end{align}
which can be different from $\psi_{\lambda_1}$. While more than two scattering layers can be computed, two layers are sufficient for most practical signal classification tasks. In order to achieve translation invariance in the scattering coefficients, the outputs of each layer are averaged in time by correlation with a low pass filter $\phi$, whose temporal support is given by the time $T$. The vector of coefficients $\V S$ is composed of the time averaged outputs of each scattering layer, resulting in a hierarchical structure with an increasing number of coefficients per layer.\\
In this paper, the Python library Kymatio \cite{Andreux2022} is used to compute the scattering transform. Apart from the parameters $Q$ and $T$, which define the frequency resolution of the wavelet basis and the length of the averaging low pass filter, the maximum scale of the wavelet basis $2^J$, with $J \in N$, can also be specified. In this software environment, the mother wavelet shape is fixed to complex valued Morlet Wavelets.\\

\subsection{Prinicpal component analysis} \label{sec:PCA}
Principal component analysis (PCA) \cite{Jolliffe2002} is an orthogonal, linear coordinate transformation, which describes a dataset by a new basis, ordered by variance. It is commonly used for dimensionality reduction and has become a common tool in data science and engineering.\\
In the context of PCA, a dataset is given by a data matrix $\M D$, in which separate data samples $\V d$ are arranged in rows, while the columns contain the feature values corresponding to each data sample. The aim of principal component analysis is to represent $\M D$ in the principal component space $\M H$, where the first principal axis lies in the direction of highest variance in $\M D$. For each subsequent principal axis, their variance is maximized under the condition that it is orthogonal to all previous principal axes.
This is achieved by applying an orthogonal transformation $\M U$, so that
\begin{align}
	\M H = \M D \M U.
\end{align}
It can be shown that the columns of the transformation matrix $\M U$ are the eigenvectors $\V u$ of $\M D^T \M D$, ordered by the size of their eigenvalues. These eigenvectors $\V u$ can also be obtained via a singular value decomposition of the data matrix $\M D$, which is the standard way of computing $\M U$.\\
Transforming a data sample $\V d$ into principal component space
\begin{align}
	\V h = \V d \M U
\end{align}
simply describes it as a weighted sum of eigenvectors $\V u$, where the weights $\V h$ of these eigenvectors are called principal component (PC) scores. Given the PC scores $\V h$, the original data sample can be reconstructed
\begin{align}
	\V d = \V h \M U^T
\end{align}
since $\M U$ is orthogonal. A dimensionality reduction can be achieved by simply truncating the transformation matrix
\begin{align}
	\M H_\text{red} = \M D \M U_\text{red}
\end{align}
and discarding the PC axes below a selected variance threshold $\theta$. Because the principal axes in $\M H$ are ordered by variance, in most examples using real world data, the first eigenvectors $\V u_{1,...,\theta-1}$ contribute significantly more to the reconstruction of $\V d$ than the discarded low variance eigenvectors. Using this reduced transform, a datapoint $d$ can be represented by the PC scores of the truncated basis
\begin{align}
	\V h_\text{red} = \V d \M U_\text{red},
\end{align}
which is described as dimensionality reduction, since the number of elements in $\V h_\text{red}$ can be significantly lower than the dimensionality of $d$, while still achieving a low reconstruction error
\begin{align}
	\V d_\text{red} = \V h_\text{red} \M U_\text{red}^T \approx \V d
\end{align}
depending on the particular dataset and the chosen threshold $\theta$.\\
PCA computations in this work are carried out using the Python library Scikit-Learn \cite{Pedregosa2011}. Since an important requirement of PCA is that each feature has a mean of zero, the feature-wise mean over all samples in $\M D$ is computed and subtracted before the PCA is applied.

\section{Experiments} \label{sec:experiments}
This paper demonstrates novel fatigue monitoring methods, which are parameterized and validated using a dataset of eBike measurements. In order to facilitate further research, this dataset is made publicly available \cite{Heindel2022} via the DOI \href{http://dx.doi.org/10.25532/OPARA-189}{\textit{http://dx.doi.org/10.25532/OPARA-189}}.

\subsection{Experimental setup and data} \label{sec:experimental_setup}
\begin{figure}
	\begin{center}
		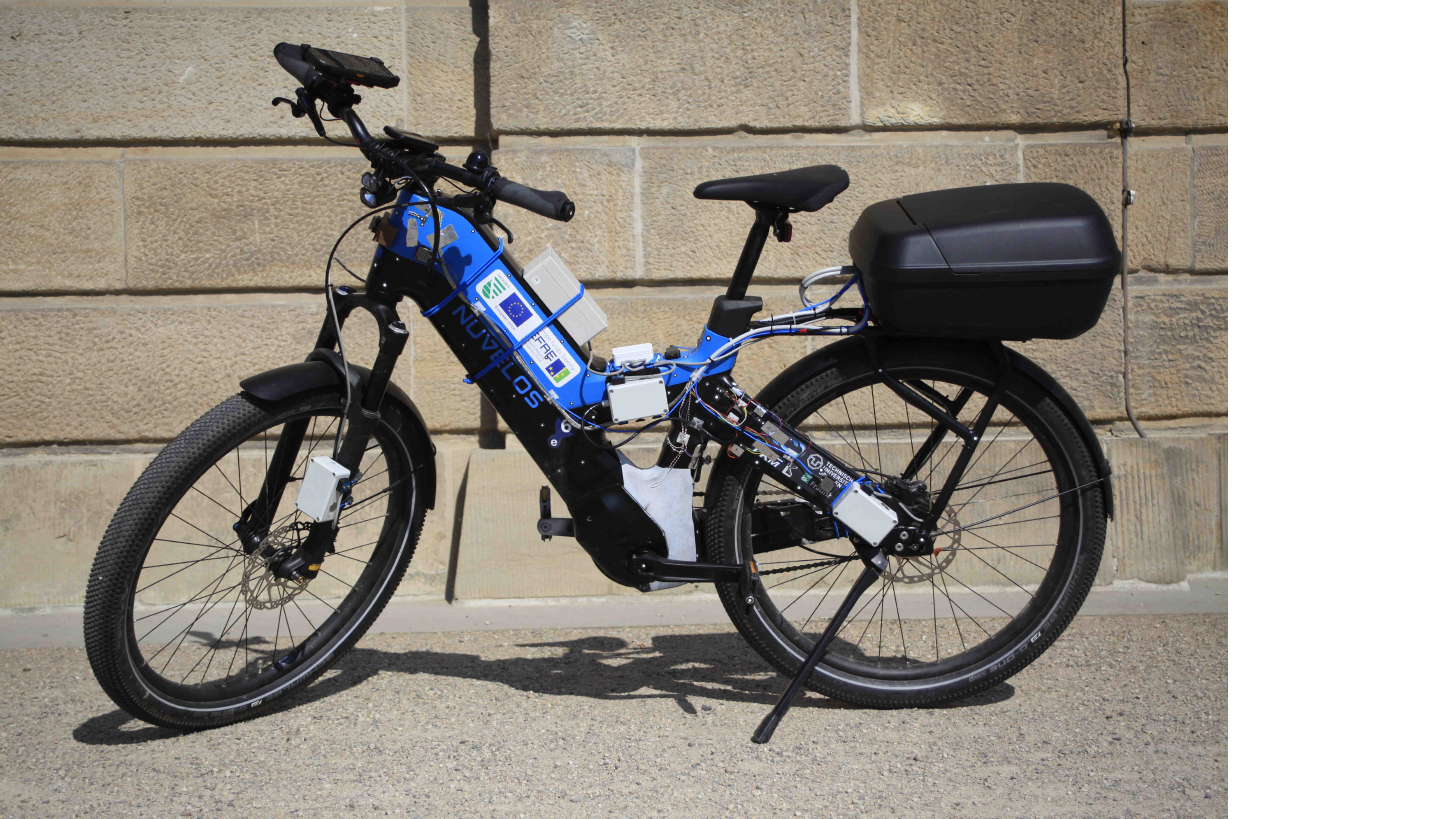\\ 
		\vspace{0.3cm}
		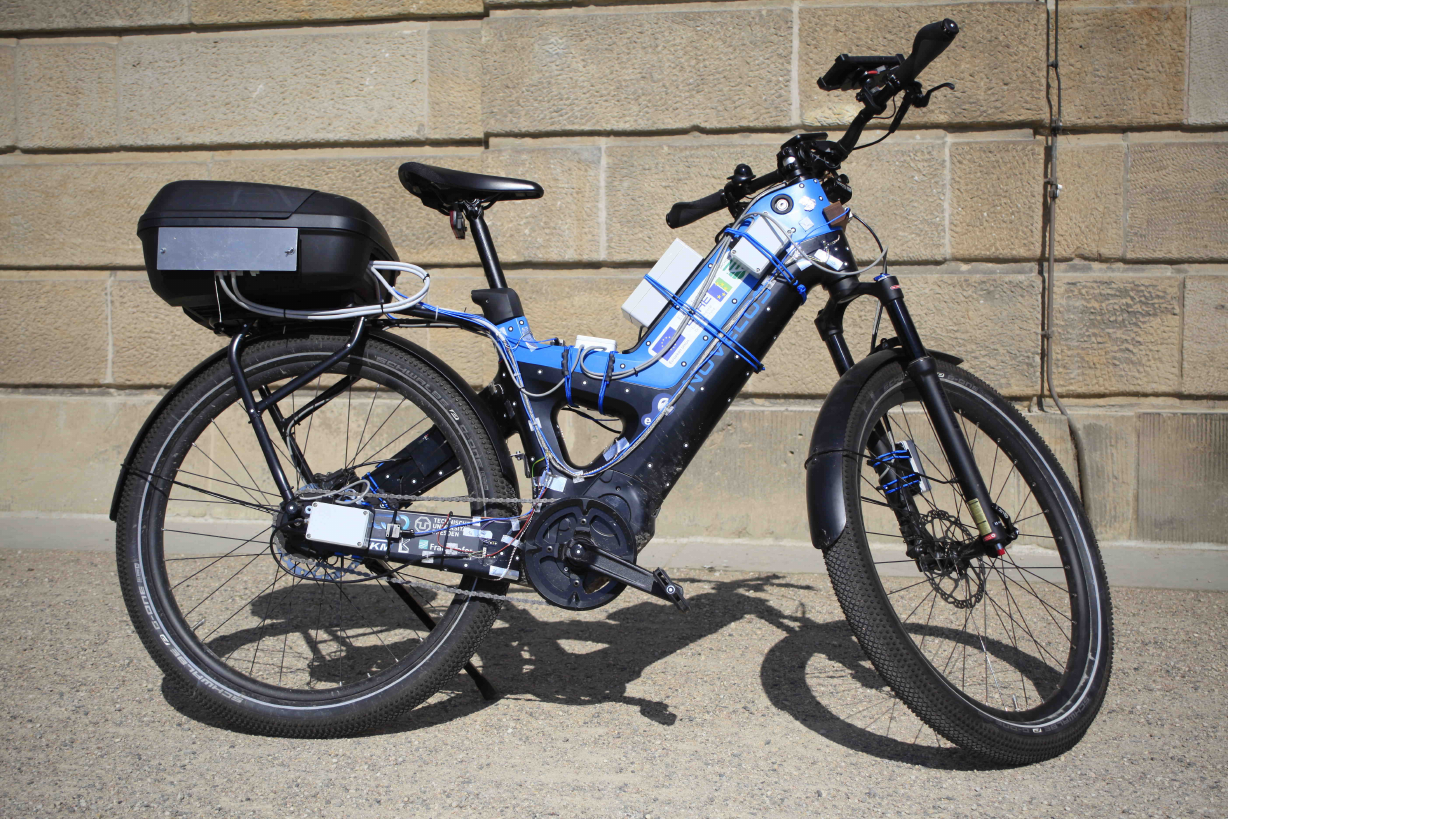 
		\caption[eBike detailed]{The measurement eBike is equipped with a variety of sensors. Their detailed locations are marked with "A" for acceleration sensors and "S" for strain gauges. The measurement amplifier is labeled as "MA", while "TC" refers to the temperature compensation gauges, which are attached to the side of the measurement box on a piece of fibre reinforced Bike frame material. The white boxes contain a different sensor setup which is not used in this study.}
		\label{fig:eBike_detailed}
	\end{center}
\end{figure}

\begin{figure}
	\begin{center}
		\subfloat[Acceleration sensor]{
			\includegraphics[height=0.34\textwidth]{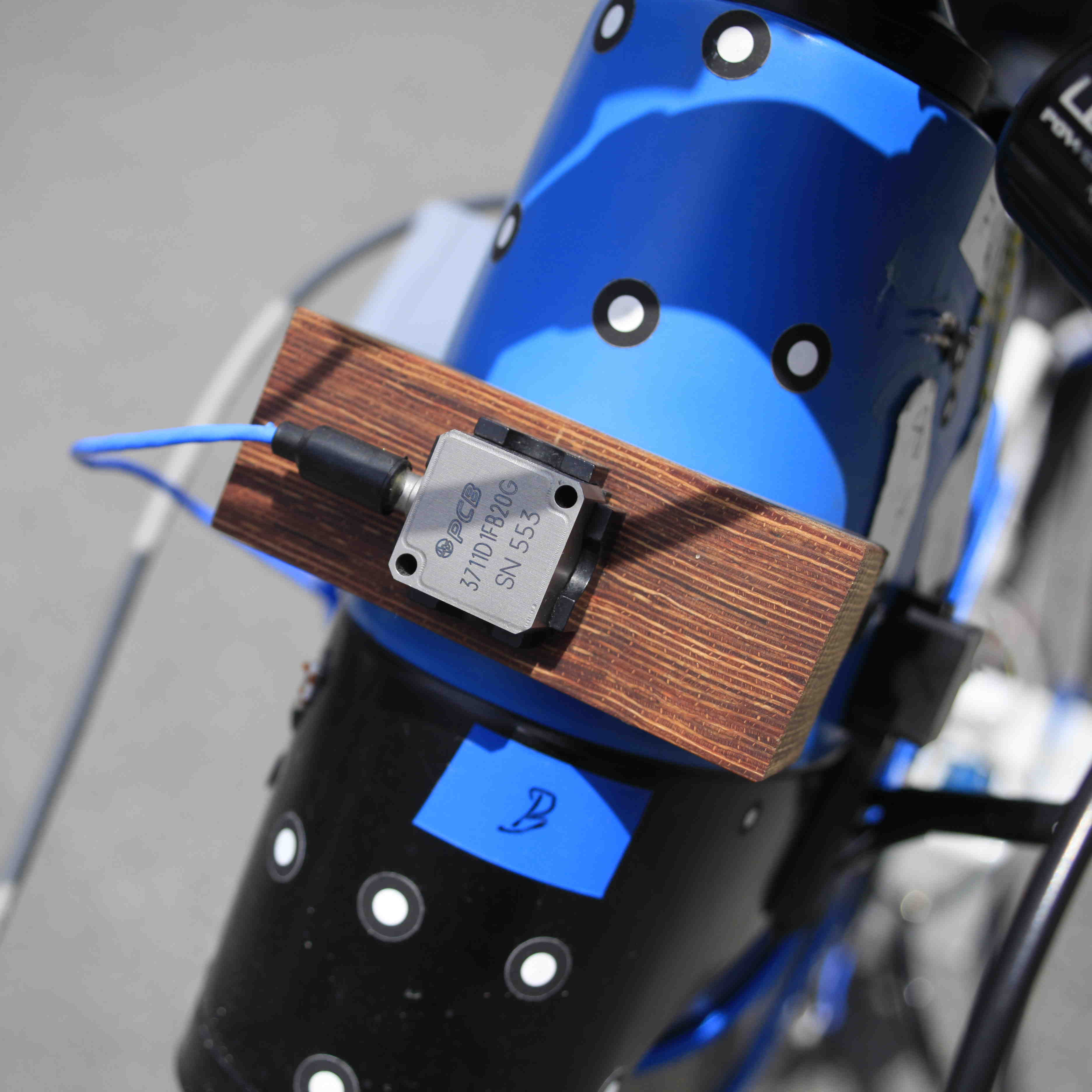}
		}
		\subfloat[Strain gauge]{
			\includegraphics[height=0.34\textwidth]{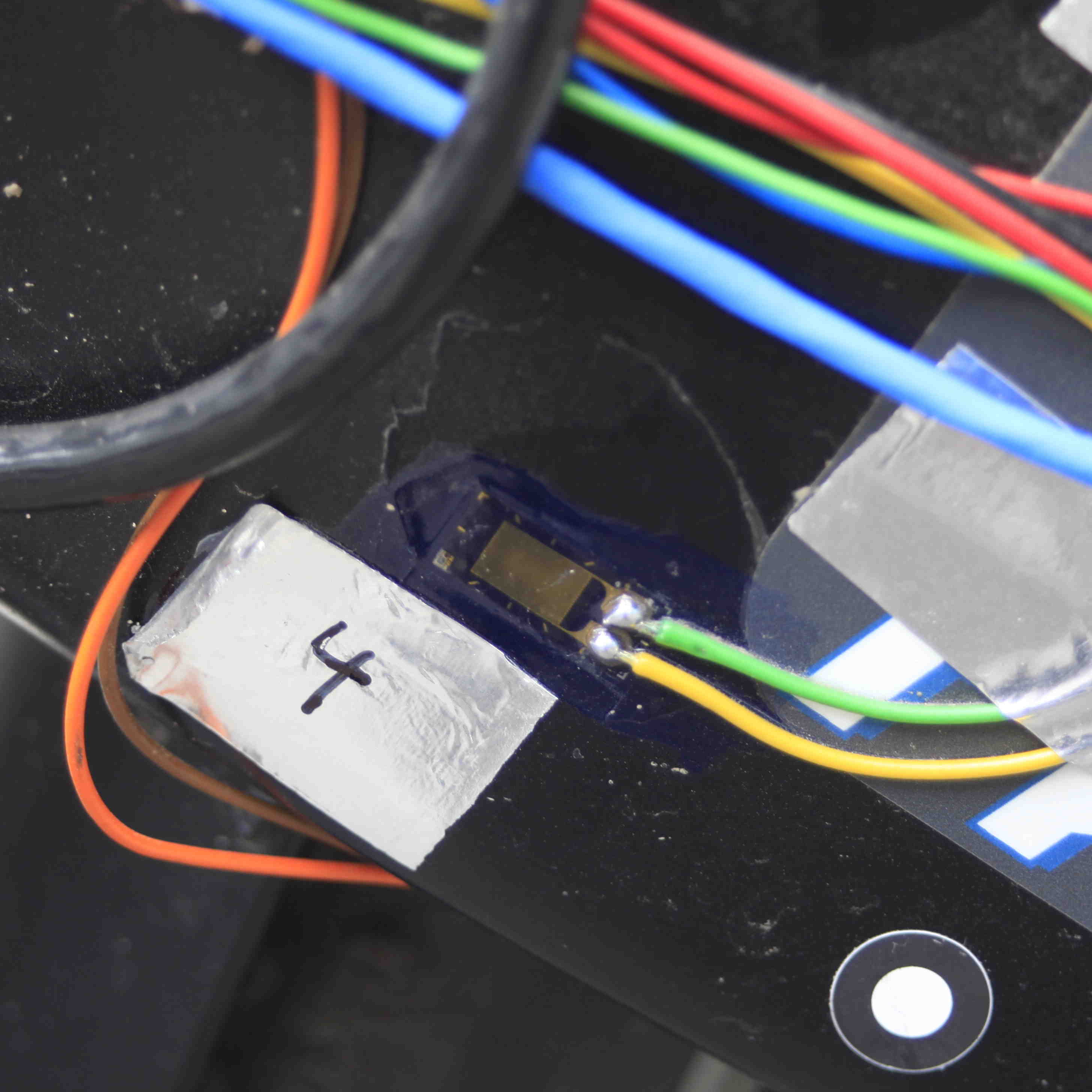}
		}
		\caption[sensors]{The sensor setup contains a total of 5 acceleration sensors (a) and 16 strain gauges (b).}
		\label{fig:sensors}
	\end{center}
\end{figure}

Experimental data is collected using the instrumented eBike of the brand Nuvelos, depicted in \autoref{fig:eBike_detailed}. Its frame is equipped with 5 uniaxial acceleration sensors and 16 strain gauges. One sensor of each type is shown exemplarily in \autoref{fig:sensors}.\\ 
While riding, the sensor data of all acceleration and strain channels is processed synchronously using two measurement amplifiers, located in the box on the rear rack. The data is then forwarded to a laptop and recorded with a sampling frequency of \SI{1200}{\hertz}. This comparatively high sampling rate was chosen in order to accurately resolve peaks in the strain and acceleration measurements. A comparison to an artificially decreased sampling rate of \SI{600}{\hertz} yields overall similar results with slightly worse performance across most evaluation metrics. The only preprocessing step consists in removing data at the start and end of each file where the measurement is being started, but the eBike is not yet in motion.\\
During the conducted measurement rides and maneuvers, the average strain amplitudes varied significantly between strain sensor locations. In order to keep the presented results clear and compact, sensor locations which exhibited an absolute average strain below a threshold of \SI{150}{\micro\meter/\meter} are omitted in the following investigations, as these locations are assumed to be less critical regarding fatigue. This comparison between strain channels was conducted using data from rider 1, riding at \SI{15}{\kilo\meter/\hour} on a cobblestone underground.\\

\subsubsection{Unlabeled usage data}
The unlabeled dataset includes regular eBike usage data from three different riders. For each individual data file, only the corresponding rider is known, but information on the precise order of events during the ride and the exact route are not available. It features various ordinary driving undergrounds like roads and bike paths, as well as park and forest trails. Since medium to high electric assistance of the eBike was used during most of the rides, the riding speed is mostly in the range of \SI{20}{\kilo\meter/\hour} to \SI{25}{\kilo\meter/\hour}. The respective weights of the individual riders are \SI{66}{\kilogram} for rider 1, \SI{83}{\kilogram} for rider 2 and \SI{110}{\kilogram} for rider 3.\\
This dataset contains 26 files with an individual length between 3 and 32 minutes, totaling 5.1 hours of unlabeled data.

\subsubsection{Labeled data}
\begin{figure}
	\begin{center}
		\subfloat[Even training]{
			\shortstack{
				\includegraphics[height=0.22\textwidth]{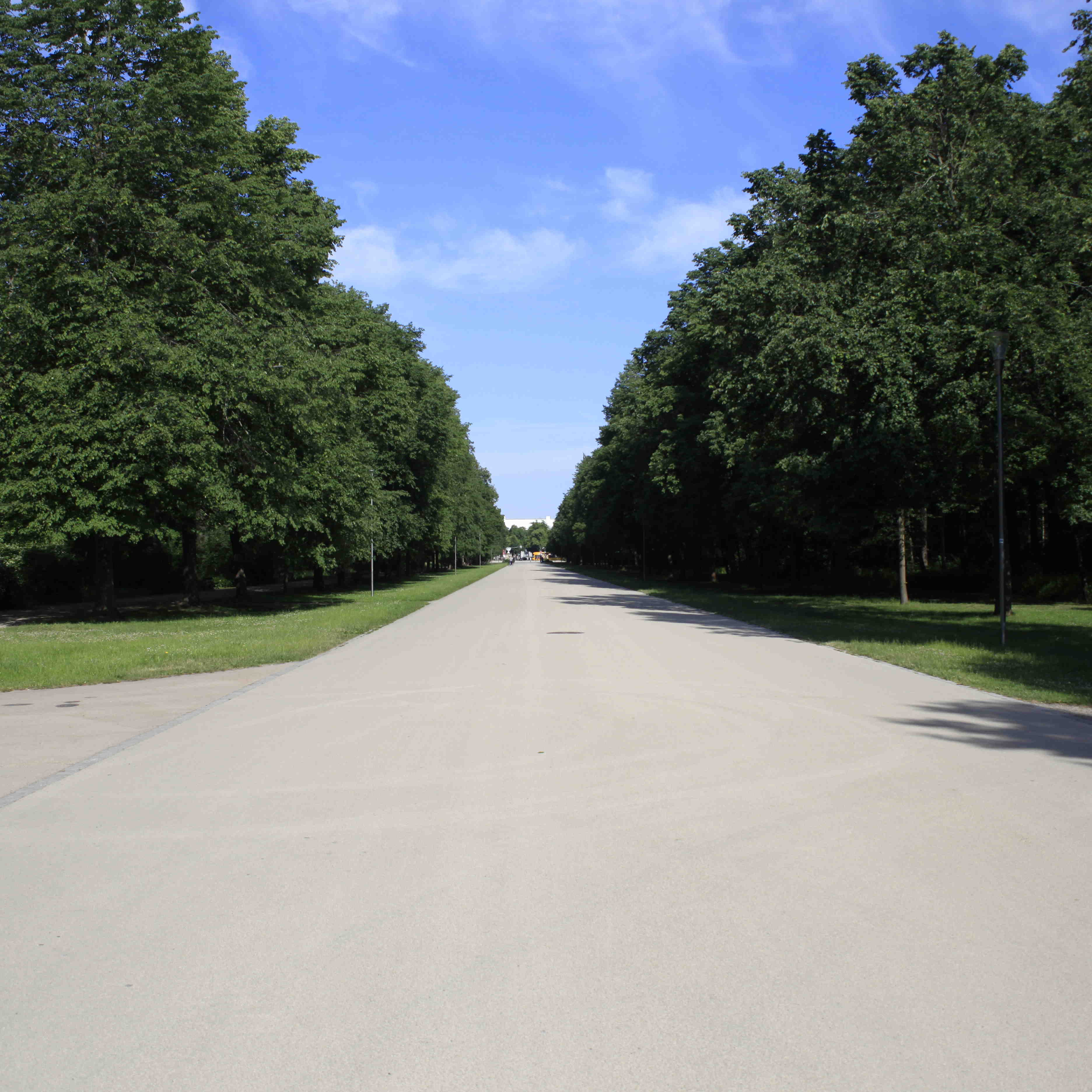}\vspace{0.005\textwidth}\\            
				\includegraphics[height=0.22\textwidth]{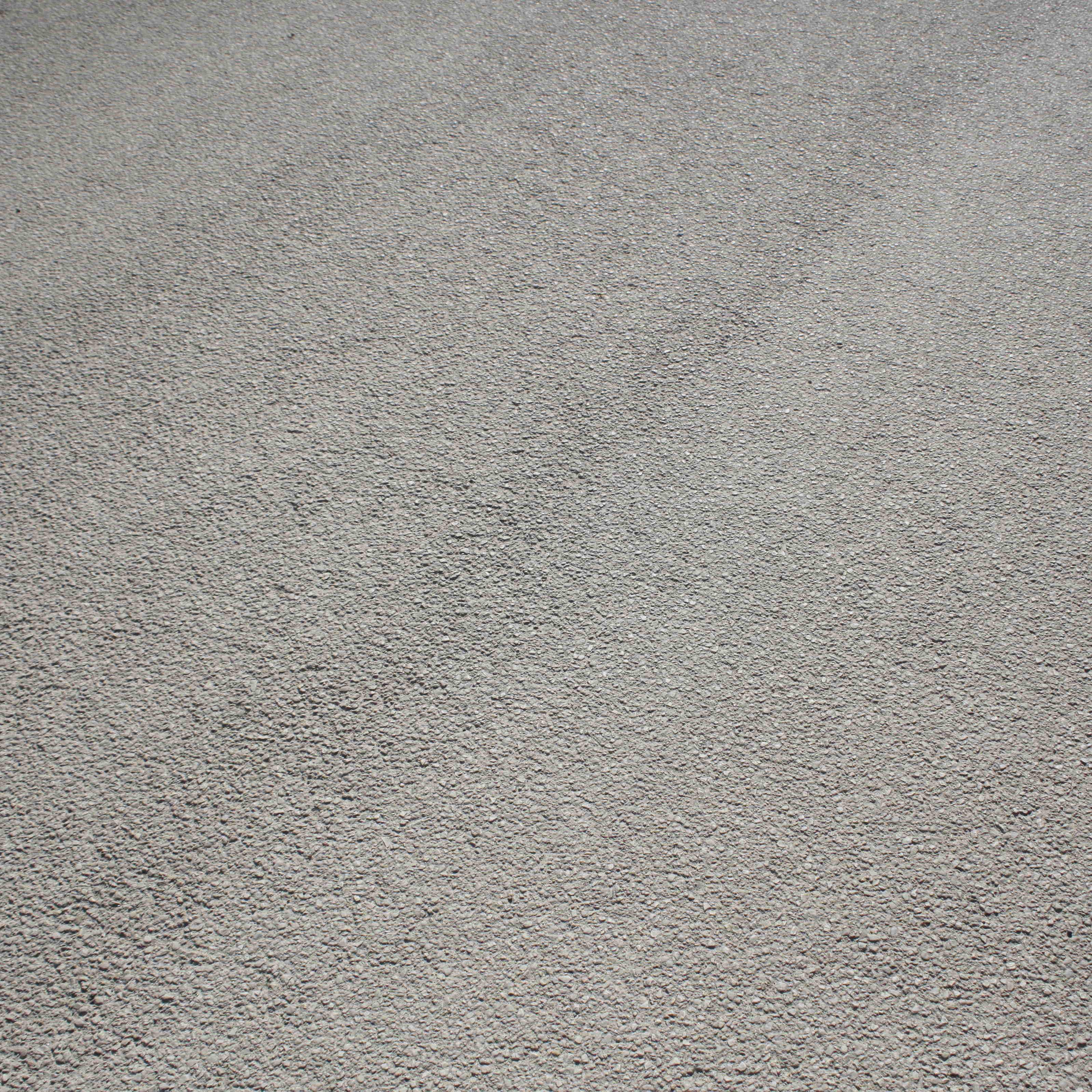}}		
		}
		\subfloat[Even testing]{
			\shortstack{
				\includegraphics[height=0.22\textwidth]{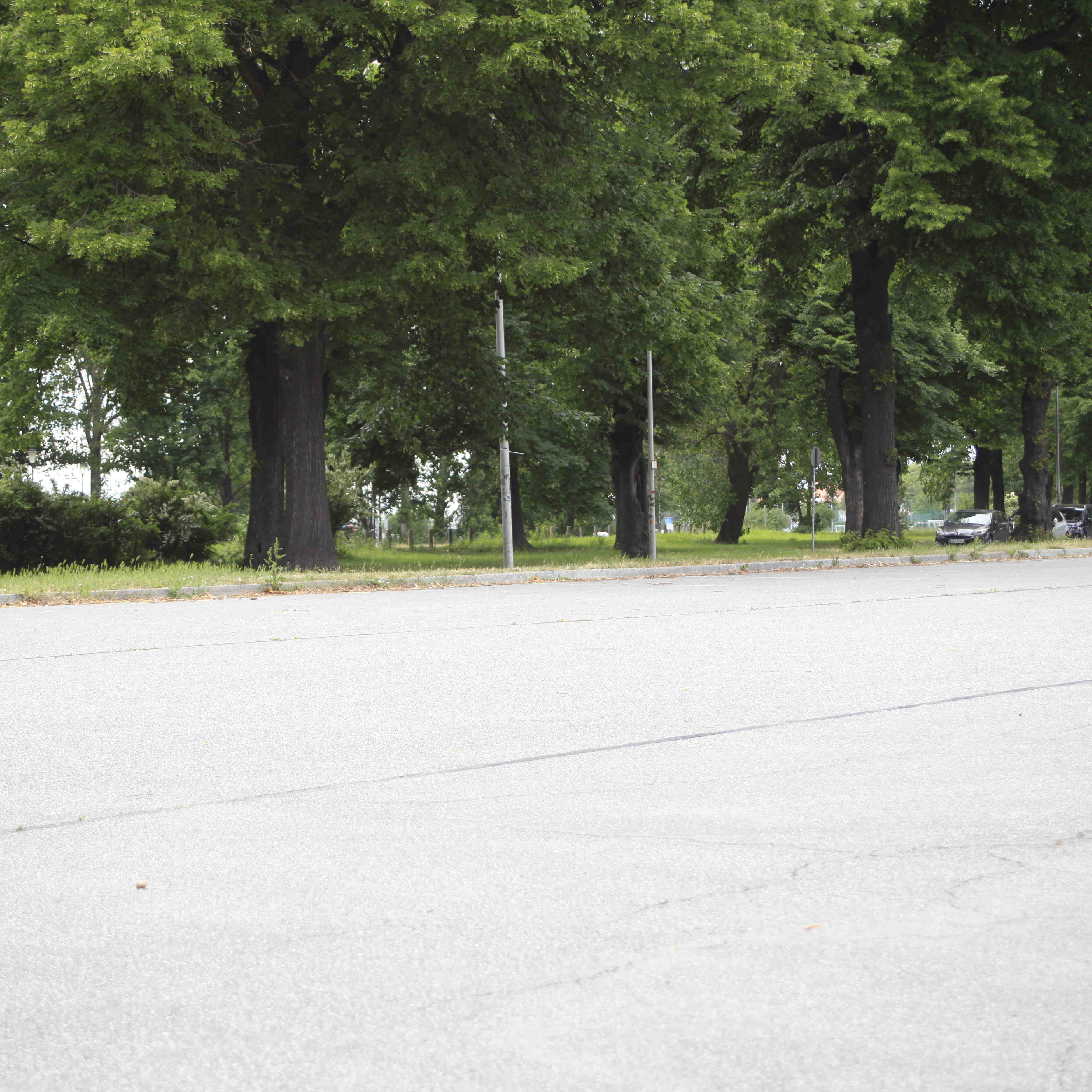}\vspace{0.005\textwidth}\\            
				\includegraphics[height=0.22\textwidth]{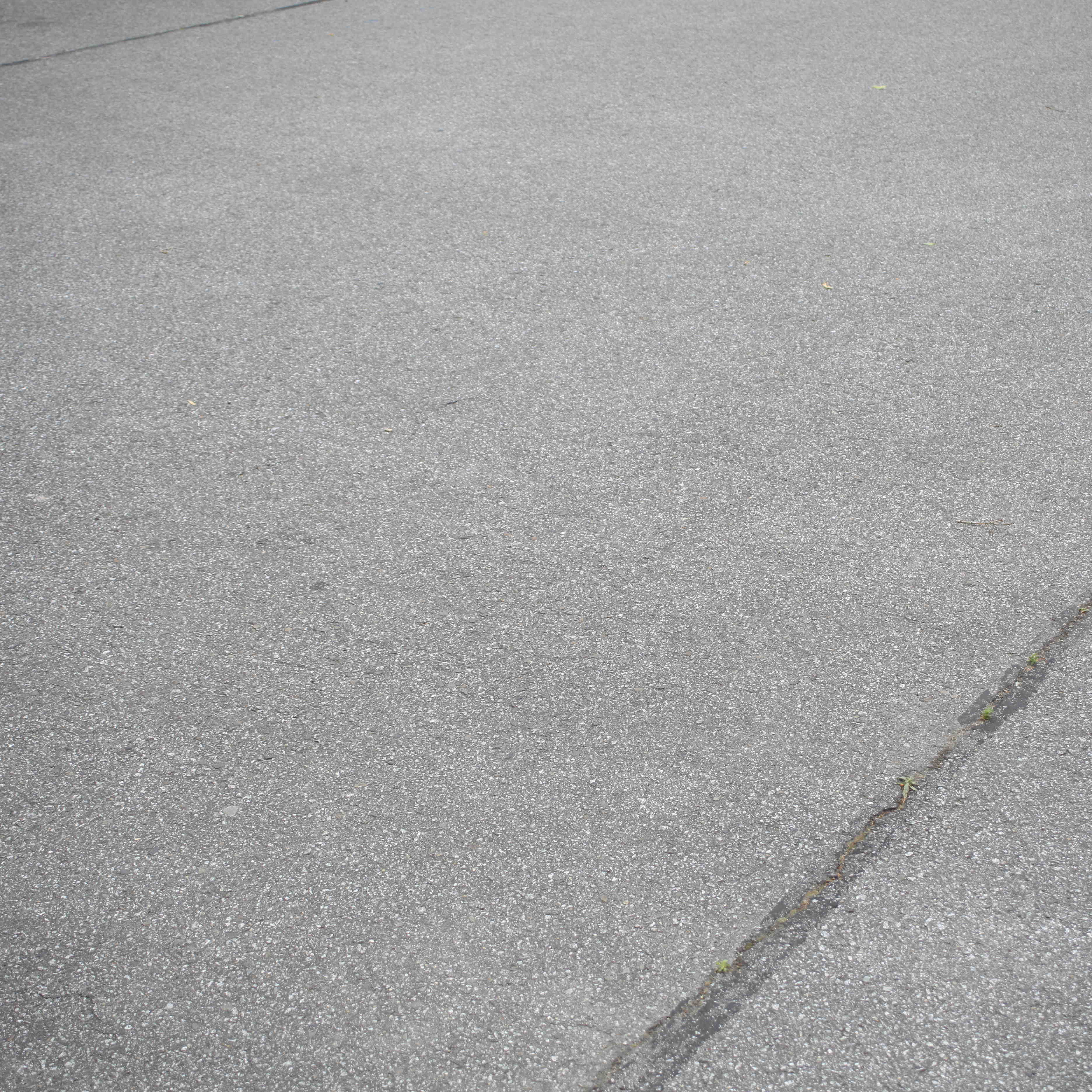}}
		}
		\subfloat[Cobble training]{
			\shortstack{
				\includegraphics[height=0.22\textwidth]{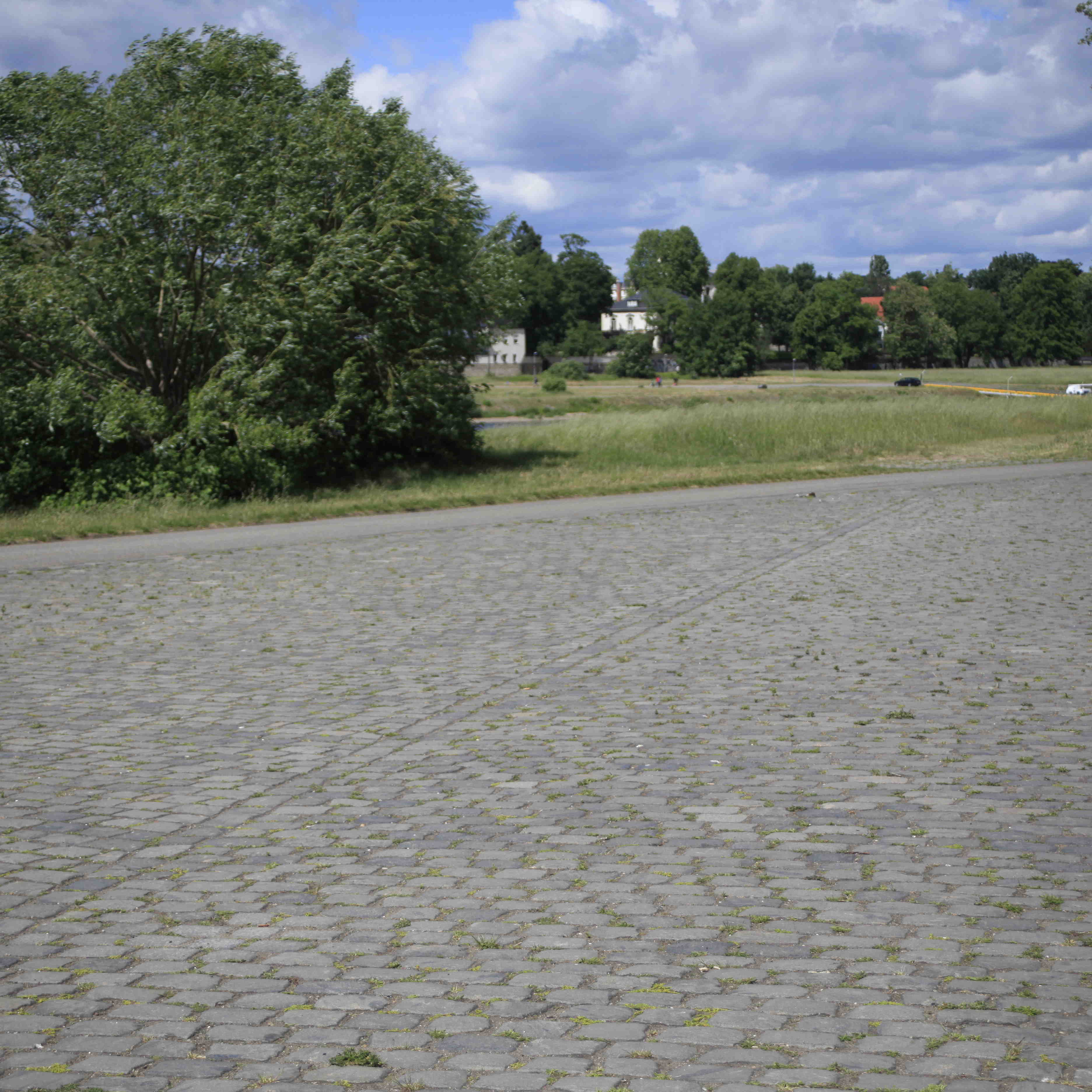}\vspace{0.005\textwidth}\\
				\includegraphics[height=0.22\textwidth]{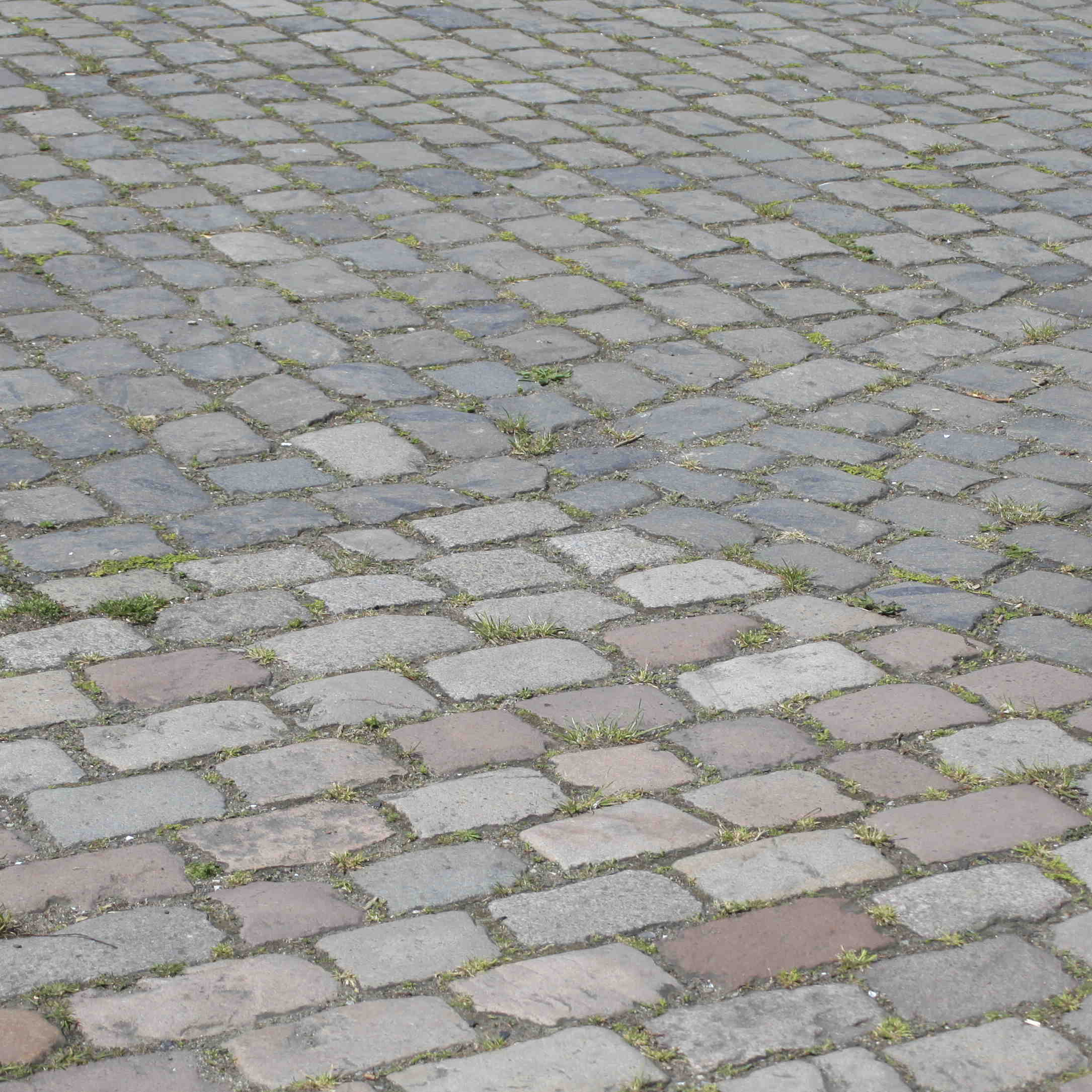}}
		}
		\subfloat[Cobble testing]{
			\shortstack{
				\includegraphics[height=0.22\textwidth]{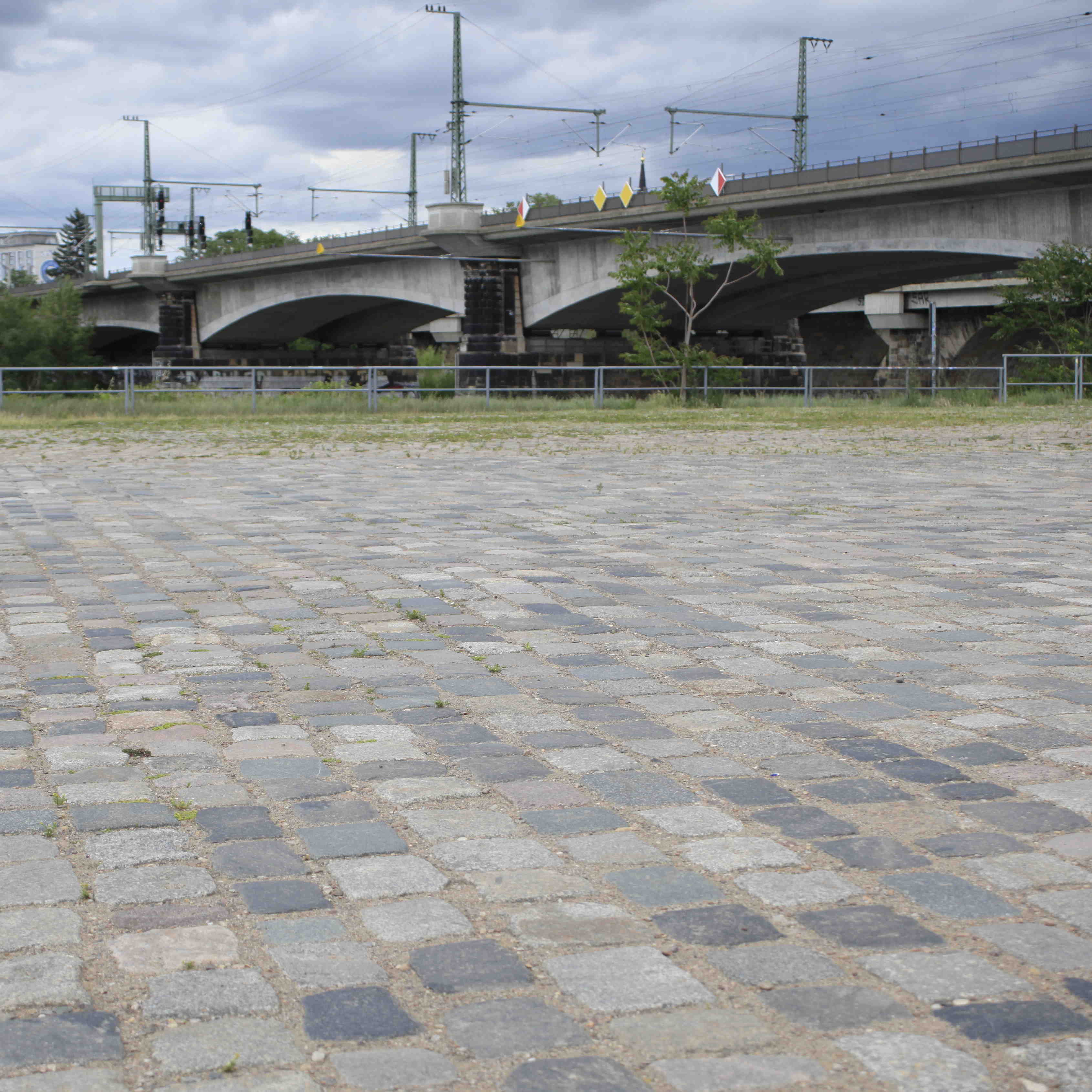}\vspace{0.005\textwidth}\\            
				\includegraphics[height=0.22\textwidth]{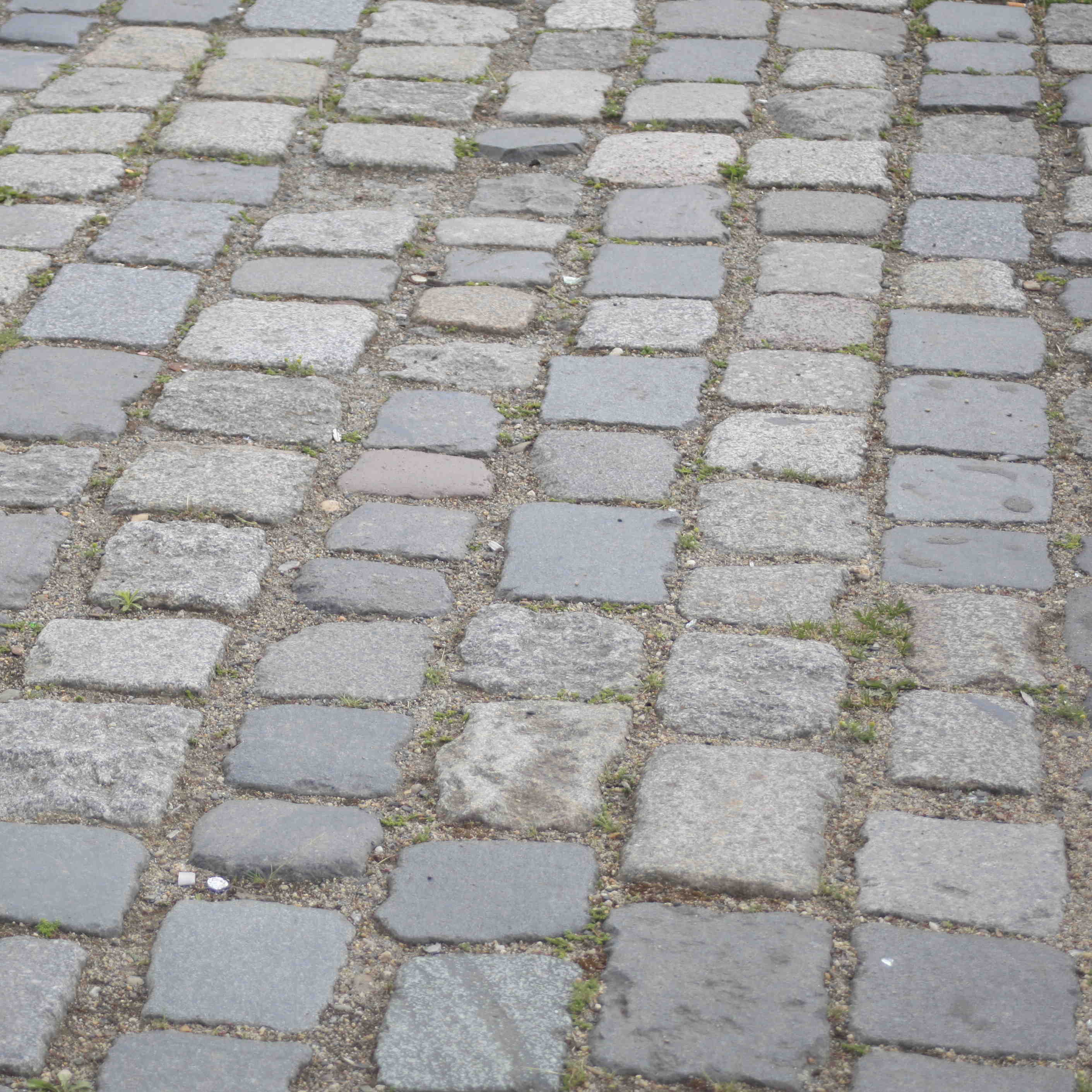}}
		}
		\caption[sensors]{Labeled measurements on even underground where collected in location (a) and (b), while uneven cobblestone underground measurements were conducted at locations (c) and (d). In order to avoid dependencies in the sensor data between training (a,c) and testing (b,d) different measurement sites were used for the same underground type.}
		\label{fig:underground}
	\end{center}
\end{figure}

In addition to the general eBike usage data, specific driving situations where measured. In this second, smaller dataset each measurement file only contains data on one specific underground with a specified velocity. Information about the rider, the underground and the target velocity is available as labels to the file. While the riding speed was controlled manually during these measurement rides, the respective target speed could still be maintained within a limit of $\pm \SI{1.5}{\kilo\meter/\hour}$.\\
The labeled dataset includes an even asphalt underground and a very uneven cobblestone surface. For both undergrounds, measurements where taken in two different locations in order to provide independent data for model parameterization and testing purposes. For the first asphalt location, depicted in \autoref{fig:underground} a), measurements at 5, 10, 15, 20 and \SI{25}{\kilo\meter/\hour} are available. The second asphalt location, see \autoref{fig:underground} b), was slightly more restricted, therefore only measurements from 5 to \SI{20}{\kilo\meter/\hour} where collected. For both cobblestone measurement sites, depicted in \autoref{fig:underground} c) and d), measurements at 5, 10 and \SI{15}{\kilo\meter/\hour} are available. This measurement program was collected in identical fashion for both rider 1 and rider 2, while no labeled data is available for rider 3.\\

\subsubsection{Data processing workflow}
\begin{figure}
	\begin{center}
		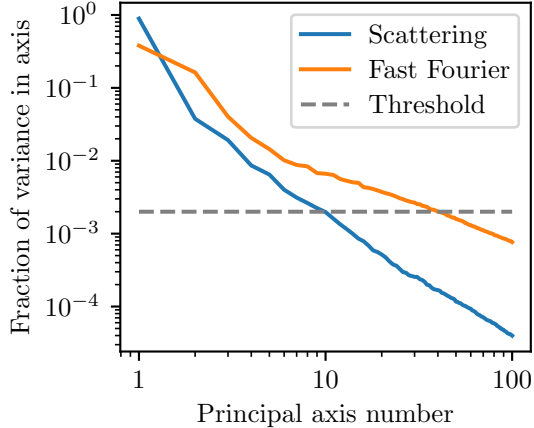
		\caption[PC score variance cutoff]{The variance content in each principal axis is steadily decreasing with each additional component. The variance content in the first principal axis is higher for the scattering based approach, followed by a steep descent. In order to achieve a dimensionality reduction, a cutoff was performed at \SI{0.2}{\%} of total variance per component.}
		\label{fig:variance_cutoff}
	\end{center}
\end{figure}

The data processing workflow provided by this work is targeted at information monitoring for predictive maintenance. Regarding a fleet of many identical vehicles, this can be efficiently achieved by selecting a single vehicle as a reference, which is equipped with acceleration and strain sensors. Both types of sensors are required for model parameterization, since a relationship between acceleration and strain data has to be established. During model deployment, predictions are made based on only acceleration sensors, leading to a reduced sensor setup for all remaining standard vehicles. The sensor equipped eBike, introduced in \autoref{sec:experimental_setup}, serves as the demonstrator for this approach.\\
The overall structure of the workflow consists of time series segmentation, the application of a data transformation and lastly, dimensionality reduction. While the scattering transform is specifically designed to extract features from signal which are relevant for classification, it can be substituted with different data transforms, according to the problem at hand. In order to compare the scattering transform to an established standard, most application examples are repeated using the Fast Fourier transform, which is generally easier to implement and interpret.\\
\begin{figure}
	\begin{center}
		\input{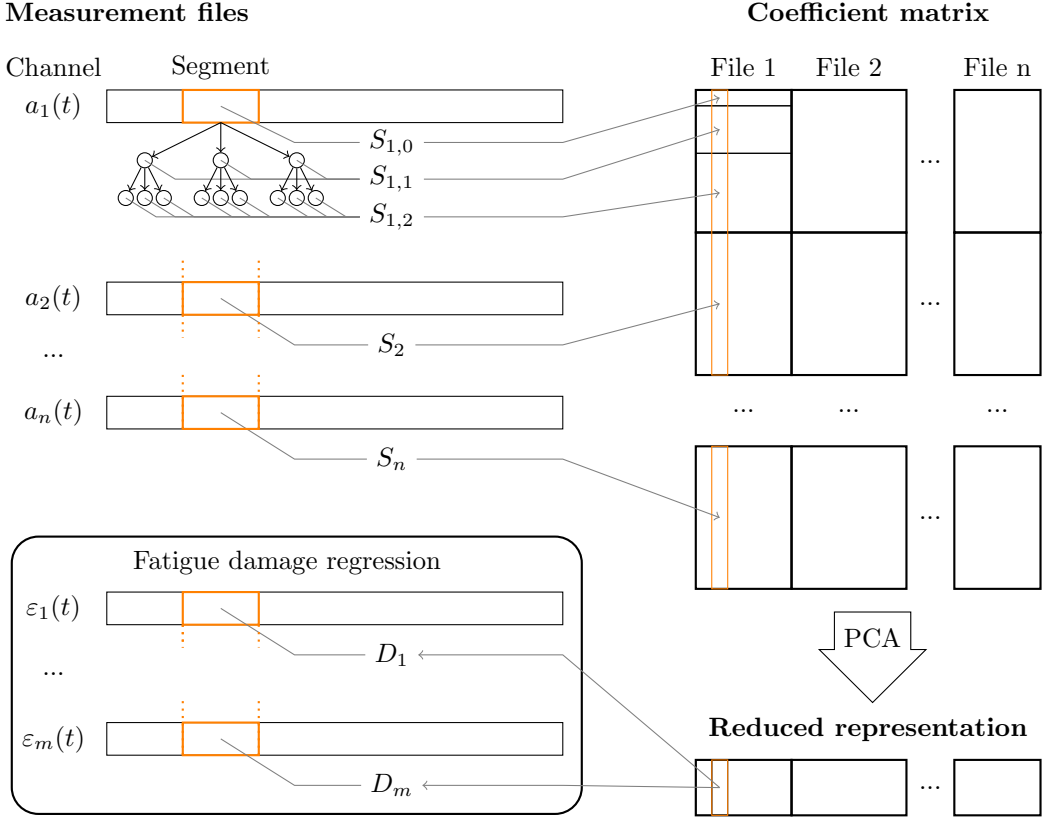}
		\caption[scattering]{Each measurement file is composed of acceleration sensor channels $a_i(t)$ and strain gauge channels $\varepsilon_j(t)$. Inside a single file, the time series data is processed in sequences. From the acceleration measurements, scattering or Fourier coefficients are computed and stored in the coefficient matrix, which is displayed in transposed form for illustrative purposes. Principal component analysis is then used to transfer the coefficient information to a reduced, low dimensional  representation, where each segment is now described by a limited set of principal component scores. This reduced representation provides the basis for fatigue monitoring applications. It can be used to directly predict fatigue damage sums $D_j$ for corresponding segments of the strain channels $\varepsilon_j$ or to predict maneuver categories using the labeled dataset.}
		\label{fig:data_processing}
	\end{center}
\end{figure}
The entire data processing workflow is depicted in \autoref{fig:data_processing}. Experimental acceleration data is obtained in the form of $n_\text{ch}=5$ channels of measurement time series with uniform sampling rate. The data is first split into shorter, non overlapping sequences called data samples, which each contain $l_\text{seq}=4096$ data points. Each data sample spans a duration of about \SI{3.4}{\second}, given the sampling rate of \SI{1200}{\hertz}. Afterwards, the chosen data transformation is applied individually to each time series segment and each measurement channel. For the scattering transform, a scale exponent of $J=5$ and a number of octaves $Q=6$ are used, which were selected through preliminary test cases. The input sequences of the Fast Fourier transform are each multiplied with a Hamming window of equal length in order to avoid discontinuities at the start and end of each sample. After computing the transformation, the coefficients of all measurement channels corresponding to one data sample are assembled in a coefficient vector. After all measurement files are processed, the complete processed data is organized in a coefficient matrix, where the columns contain the individual transformation coefficients of all channels and each rows represents a data sample from a time series segment.\\
This data matrix will be used to parameterize the principal component analysis algorithm in order to obtain a dimensionality reduction. As preprocessing, the data is first standardized seperately in each channel and in the case of the scattering transform, individually for each scattering layer. Then, the mean of the data matrix is subtracted and the PCA is applied. This requires deciding how many principal axes to keep in the truncated basis $\V h_\text{red}$ by choosing a suitable variance cutoff $\theta$. As shown in \autoref{fig:variance_cutoff}, a cutoff of at least \SI{0.2}{\%} of the total variance per axis was chosen, leading to 9 principal axes in the case of the scattering transform and 40 principal axes for the FFT based model. Compared to the full basis, the remaining axes amount to \SI{97.7}{\%} of the original variance for the scattering transform and \SI{76.2}{\%} for the Fast Fourier approach. While the FFT based model clearly loses information between transformation and PCA, obtaining the same \SI{97.7}{\%} as the scattering variant would require keeping about 1000 principal axes. This defeats the purpose of a low dimensional representation and would cause issues with many commonly used regression and classification algorithms.\\
After applying both data transformation and PCA, each sample is now described only by a set of 9 or 40 principal component scores, depending on the transformation, as opposed to the original acceleration measurements of 5 sensor channels with 4096 data points per sample. This compression of sensor information enables fatigue damage regression and maneuver classification applications by operating directly on the principal component scores.

\section{Fatigue monitoring} \label{sec:fatigue_monitoring}
The aim of this application is to showcase how the measurements of only acceleration sensors can be used to monitor the fatigue damage accumulation at the strain sensor locations.

\subsection{Regression setup} \label{sec:regression_setup}
While a reduced data representation is only computed from the acceleration data samples, the simultaneously measured sequences of the strain gauges are used to link the reduced acceleration information to a corresponding fatigue damage. In this example application, this is achieved by a fictitious damage calculation following the nominal stress concept \cite{Haibach2002}. While stress data is not accessible directly, a proportionality of nominal stresses and local strains is assumed. The strain sequences are first rainflow cycle counted using the second pass of the 4-point algorithm \cite{McInnes2008}, leading to a rainflow matrix $\M M_\text{RF}$ and avoiding influences of a resulting residuum. In order to relate the strain hystereses in $\M M_\text{RF}$ to an endured number of load cycles $N$, a fictitious \textsc{Wöhler}-curve
\begin{align}
	N = K \cdot \varepsilon_a^{-k}
\end{align} 
with $k=5$ and $K=10^7$ is used, where $\varepsilon_a$ refers to the amplitude of the strain hysteresis. The damage accumulation is then computed according to the elementary Palmgren-Miner rule \cite{Palmgren1924}, resulting in a fictitious fatigue damage sum $D$ for each strain channel sequence.\\
\begin{figure}
	\begin{center}
		\subfloat[Unlabeled data]{
			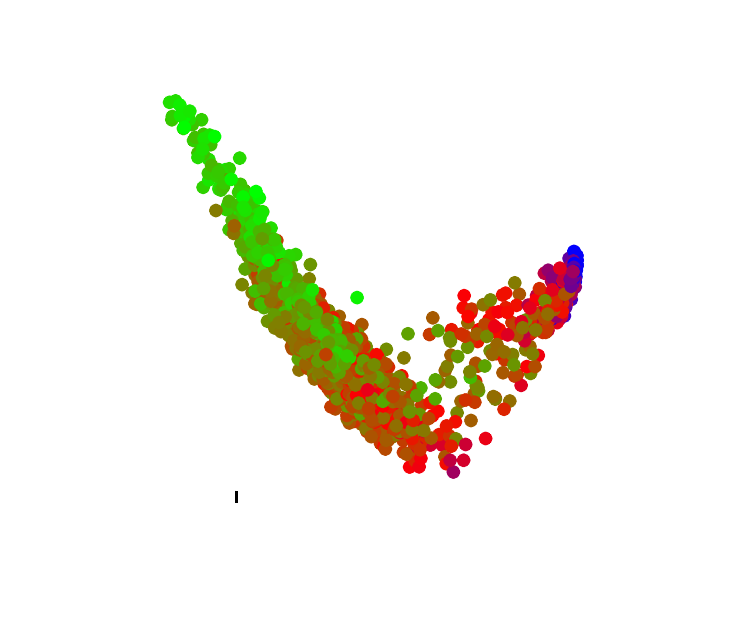
		}
		\subfloat[Labeled data]{
			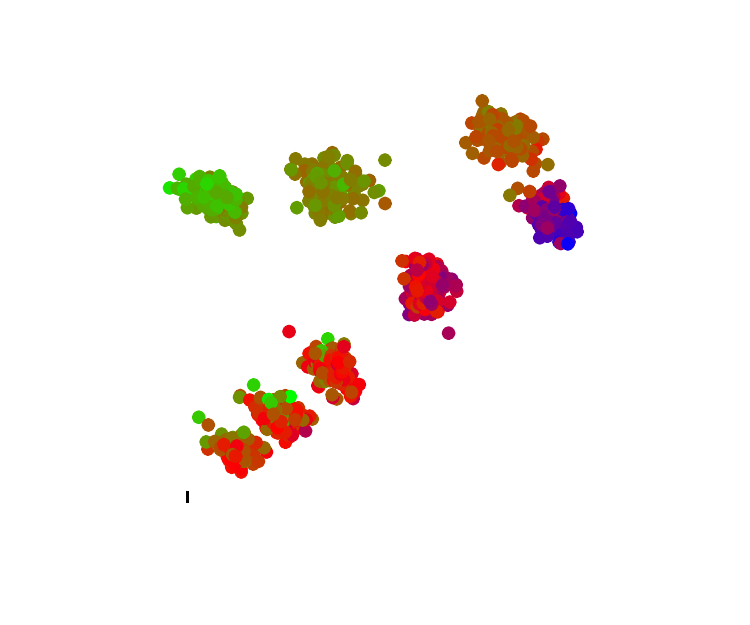
		}
		\caption[PC score vs fatigue damage]{A strong correlation exists between the principal component scores, computed from acceleration data sequences, and fatigue damage sums, stemming from concurrent measurements at strain gauge 1. Fatigue damage sums are visualized for the unlabeled training data in (a), while the labeled training data is displayed in (b). For reasons of clarity, only two selected principal axes combinations are shown.}
		\label{fig:fatigue_relationship}
	\end{center}
\end{figure}
After obtaining both the reduced representation of the acceleration data, as well as the fatigue damage sum, computed from the strain sequences, the relationship between these quantities can be visualized. In \autoref{fig:fatigue_relationship}, the fatigue damage $D$ of strain gauge number 1 is displayed as a function of the principal components. Each sequence in the unlabeled dataset corresponds to one data point in the plot, where the location of this point is given by the acceleration data and the color corresponds to the fatigue damage, computed from the concurrent strain measurements. It can be observed that a strong correlation between PC scores and fatigue damage exists, which results in local regions of low, medium or high fatigue damage. While there is a smooth transition in the damage sums of the unlabeled dataset (a), the differences between individual clusters of the labeled dataset (b) are more pronounced, see also \autoref{fig:maneuver_separation}.\\
By splitting the unlabeled data into a training- and a testing dataset, a regression model can be parameterized, which predicts the fatigue damage $D$ from the principal component scores $h_\text{red}$. In order to reduce the similarities between training and testing data, all sequences originating from a single measurement ride are grouped into one measurement file, and the train-test-split is performed on the file level. \SI{77.7}{\%} of all unlabeled samples make up the training dataset, while the remaining \SI{22.3}{\%} are used for testing. In absolute terms, this leads to a total of 4209 training samples and 1205 samples for testing. Since the rider information is available for the unlabeled data files, the distributions of riders in training and testing datasets were chosen to be very similar. Since the labeled dataset contains very uniform measurements of specific riding scenarios, it is used exclusively for testing purposes and is also omitted from the parameterization of the principal component analysis. In the regression task, the logarithmic fatigue damage sums $\text{lg} D$ are predicted, since the corresponding regression models would otherwise have to account for many different orders of magnitude in $D$. Many regression algorithms where tested, among them linear regression, polynomial regression, regression trees, lasso regression and multi layer perceptron artificial neural networks. Most of these approaches yielded a very comparable prediction performance, therefore a multivariate quadratic regression was chosen, since it is easy to implement and to interpret.\\
In order to assess the predictive performance of the regressor, two measures are taken into account. The coefficient of determination, also known as the R² error metric
\begin{align}
	R²(y,y^*) = 1- \frac{\sum_{i=1}^{n} (y_i-y_i^*)^2}{\sum_{i=1}^{n} (y_i-\bar{y})^2} \hspace{1cm} \text{with} \hspace{1cm} \bar{y}=1/n\sum_{i=1}^{n}y_i,
\end{align}
describes the fraction of the variance in the observed data $y$ which is explained by the model predictions $y^*$. It equals 1 for a perfect prediction and decreases with increasing deviations between $y$ and $y^*$. In this application, the target $y$ corresponds to the logarithmic fatigue damage sum $\text{lg} D$ obtained from measurements, while $y^*$ is the prediction prediction of the regression model, resulting from the reduced representation of acceleration measurements. Since R² is a very general regression metric, it does not differentiate between the prediction accuracy of smaller and larger fatigue damage sums. In order to capture how well the regression approach is suited for fatigue monitoring, the fatigue damage sum ratio
\begin{align}
	r_\text{FDS} = \frac{\sum_j D^*_j}{\sum_j D_j}
\end{align}
is introduced as a second metric. It compares the total sum of all $n$ non-logarithmic fatigue damage contributions $D_j$ in the testing dataset to the sum of their predictions $D_j^*$. For a perfect fatigue prediction, $r_\text{FDS}$ equals 1, while a ratio of e.g. 2 means that the true fatigue damage sum is overestimated by a factor of 2.

\subsection{Prediction of unlabeled data}
\begin{figure}
	\begin{center}
		\subfloat[Prediction scatter plot]{
			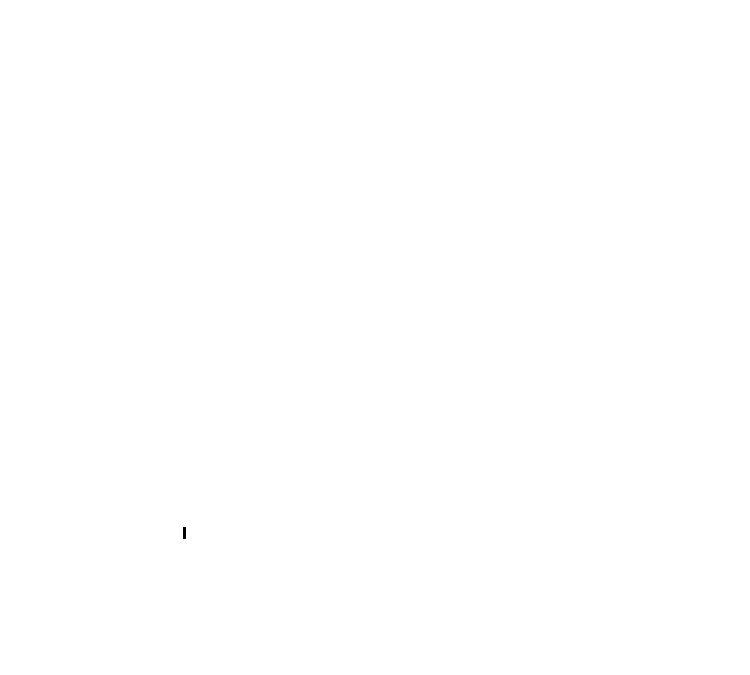
		}
		\subfloat[Error histogram]{
			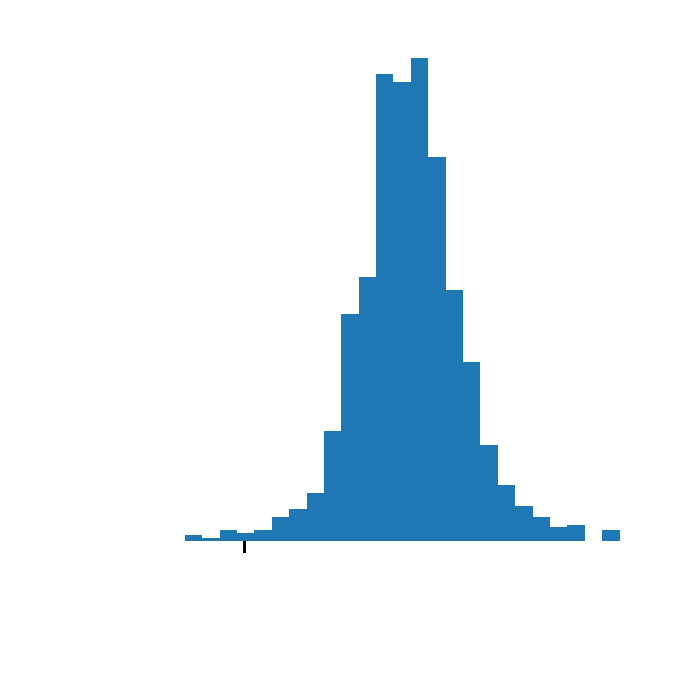
		}
		\caption[Regression example]{Fatigue damage sums $D$, computed from measurement samples at strain gauge 1, are compared to a prediction $D^*$, obtained from the low dimensional representation of concurrent acceleration measurements. All samples in the unlabeled testing dataset are shown. The grey middle line in the scatter plot (a) represents a perfect prediction, while the dashed lines each denote an error of 1 in the prediction of lg $D$, corresponding to an error of factor 10 in $D$. The prediction error (b) is slightly biased towards overprediction.}
		\label{fig:regression_example}
	\end{center}
\end{figure}

\begin{figure}
	\begin{center}
		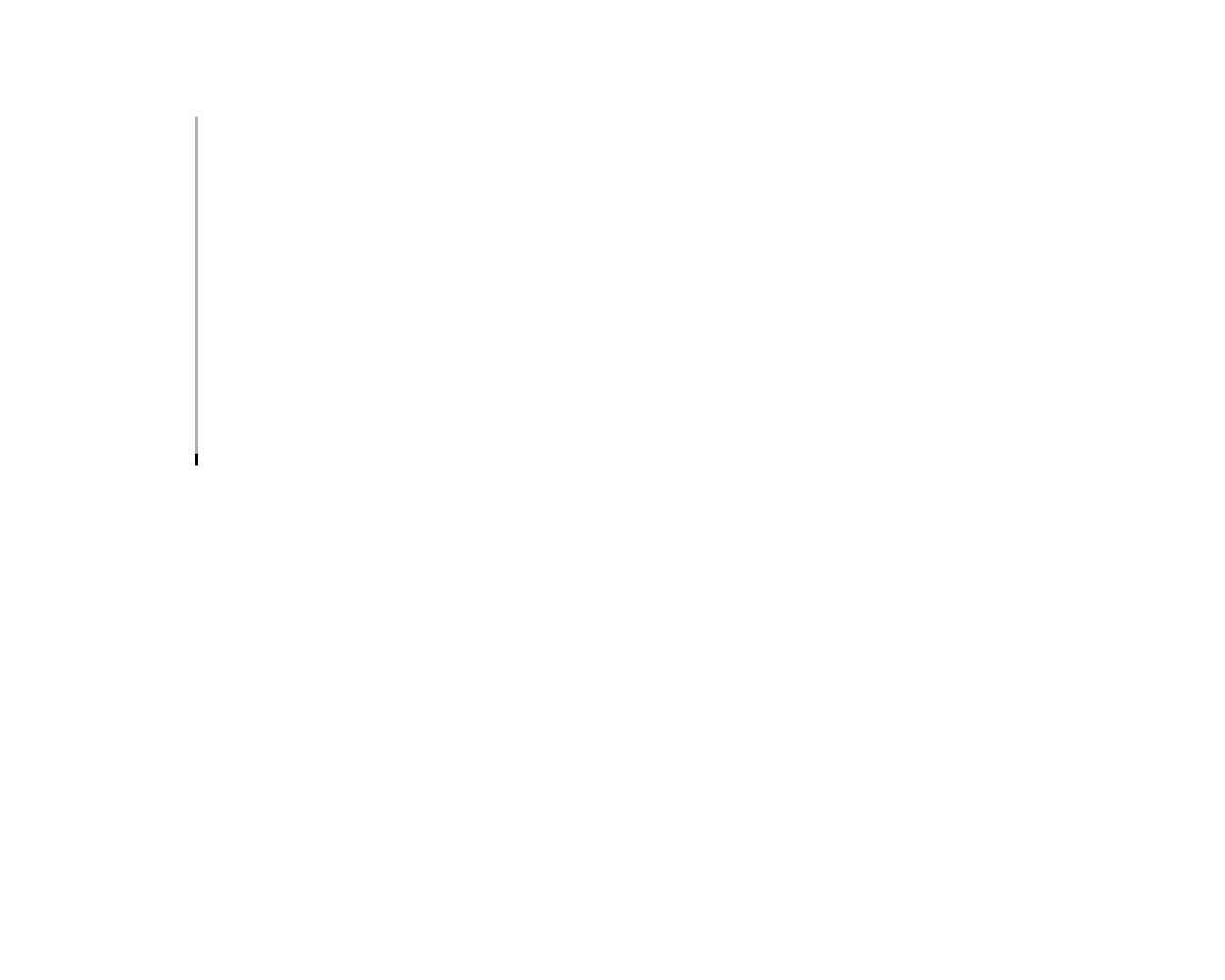
		\caption[Regression comparison unlabeled]{The low dimensional representation of acceleration sensor data is used to predict the fatigue damage sum of the corresponding strain measurements for all samples in the unlabeled testing dataset. Only strain sensors with comparatively high average amplitudes are shown.}
		\label{fig:regression_comparison}
	\end{center}
\end{figure}

Predicting the fatigue damage of unlabeled usage data is a very direct approach to fatigue monitoring. Evaluating the performance of a predictive model is increasingly difficult when many sensor channels, as well as multiple model variations are available. In order to provide a starting point for this evaluation, \autoref{fig:regression_example} exemplarily displays the application of the presented approach to the fatigue damage prediction for strain sensor 1, using the scattering transform to compute the reduced representation of acceleration data. It can be seen that the fatigue damage from measurement and prediction are very strongly correlated, indicated by an R² value of 0.83. At the same time, both the general scatter and the error histogram indicate that errors in the range of $\pm 1$ in $\text{lg} D$, i.e., one decade in $D$, are relatively common. While the mean of the error is close to 0, the predictions $D^*$ tend to be slightly larger than the computed values of $D$.\\
Given this baseline, the overall predictive performance on the unlabeled test dataset can be evaluated, which is presented in \autoref{fig:regression_comparison}. As stated in \autoref{sec:experiments}, strain sensors with very low amplitudes where omitted, as the corresponding locations are not relevant to fatigue and the measurements contain a significantly higher signal-noise-ratio. Regarding the R² plot, the regression setup using the scattering transform generally outperforms the approach based on the Fast Fourier transform significantly. This is reflected in a generally lower scatter in the prediction of individual data points, i.e., measurement sequences. Both approaches yield high overall R²-scores, averaging at roughly 0.85 for the scattering transform and 0.75 for the FFT. The fatigue damage sum ratio is generally more difficult to interpret, since high errors of individual predictions can compensate each other in the total damage sum. In this metric, the better approximation changes depending on the measurement location. Both approaches provide predictions which are mostly located in the range of $0.5 \leq r_\text{FDS} \leq 2$, meaning that the error of prediction is at most of factor 2, but the scattering based approach slightly underestimates the damage sum consistently.

\subsection{Prediction of labeled data}
\begin{figure}
	\begin{center}
		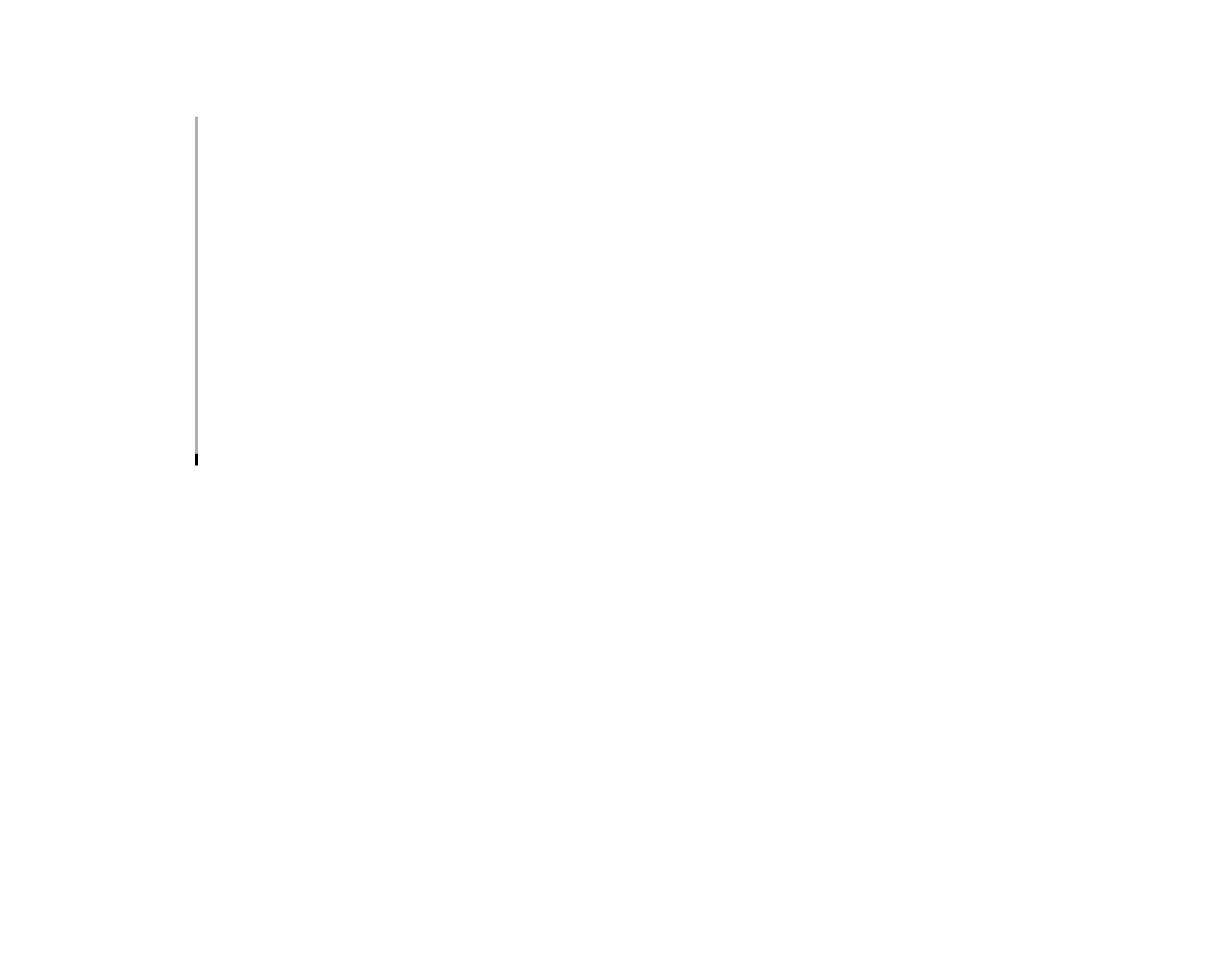
		\caption[Regression comparison labeled]{The low dimensional representation of acceleration sensor data is used to predict the fatigue damage sum of the corresponding strain measurements for all samples in the labeled testing dataset. The data samples are divided into even underground, and very uneven cobblestone underground. Only strain sensors with comparatively high average amplitudes are shown.}
		\label{fig:regression_comparison_labeled}
	\end{center}
\end{figure}

The testing measurements of the labeled dataset can be used in order to evaluate how well the fatigue damage regression performs in more extreme cases. It contains measurements taken on both very even underground, as well as very uneven cobblestone. Many of these measurements are also recorded at very low riding speeds of \SI{5}{\kilo\meter/\hour} to \SI{15}{\kilo\meter/\hour}. Since no labeled data was used to parameterize the low dimensional representation or the regression model, it showcases the regression of events which are underrepresented in the unlabeled dataset.\\
In \autoref{fig:regression_comparison_labeled}, the results of labeled data predictions are shown, grouped by even and uneven undergrounds. Compared to the unlabeled data, the general predictive accuracy decreases significantly for both scattering transform and FFT, indicated by the generally lower R² score. Especially strain sensor 11 is predicted extremely poorly by both models. Apart from this sensor, both approaches yield relatively similar results on even underground, while the scattering transform scores consistently higher results on the uneven cobblestone. This trend is also reflected in the fatigue damage sum ratio, where all scattering results are in the range of $0.5 \leq r_\text{FDS} \leq 2$, while the fatigue damage on cobblestone is overestimated consistently by the Fast Fourier based model.

\section{Maneuver identification}
In \autoref{sec:fatigue_monitoring}, a strong correlation between the reduced representation of acceleration data and the fatigue damage accumulation was shown. Maneuver identification is an alternative approach to direct fatigue damage regression, where the aim is to correctly identify the current dynamic state of the vehicle. This information can afterwards be used to associate each maneuver with a corresponding damage accumulation model, but it can also provide general usage statistics to vehicle design engineers or fleet managers. The aim of this section is to show how the proposed low dimensional data representation can be used to identify maneuvers by providing a discrete classification example.

\subsection{Semi-supervised parameterization}
\begin{figure}
	\begin{center}
		\subfloat[Underground separation]{
			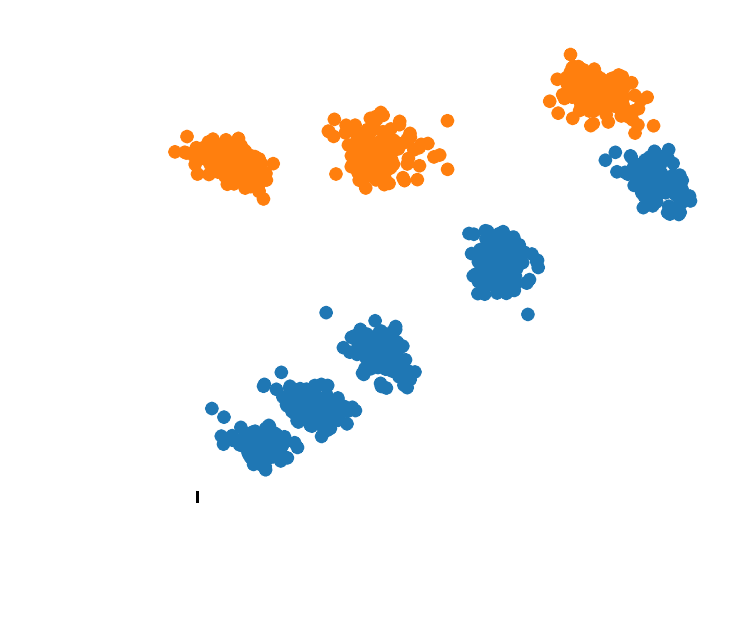
		}
		\subfloat[Riding speed separation]{
			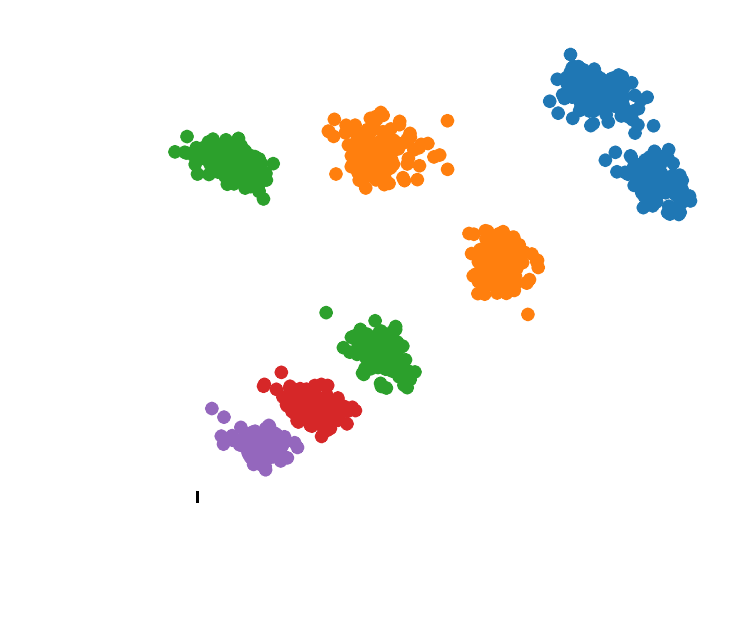
		}
		\caption[Maneuver separation]{Using the training portion of the labeled dataset, the location of specific maneuvers in the principal component space can be identified. Even and uneven undergrounds (a) can be separated mostly along PC 2, while a decrease in PC1 corresponds to an increase in riding speed (b).}
		\label{fig:maneuver_separation}
	\end{center}
\end{figure}

A main difficulty with classification tasks is the collection of labeled data examples. In order for a data-driven algorithm to identify a specific maneuver of interest, examples of this maneuver have to be provided. This process can be very cost intensive, since the corresponding data has to be either collected in targeted maneuver measurements or manually extracted from general usage data, for example by using rider logs of each measurement or video recordings. As a result, labeled data is rarely available in large quantities, which should be taken into account when developing a maneuver identification algorithm.\\
Since maneuver data should be collected from all vehicles in a fleet, only the acceleration sensors of the eBike setup, presented in \autoref{sec:experimental_setup}, are used in this study. From the acceleration data, the reduced representation is build exactly as described in \autoref{sec:regression_setup}. First, the data is split into segments, then either the scattering or Fast Fourier transform is applied and finally, the PCA is used to reduce the dimensionality. This process uses only unlabeled usage data, which is generally easy to obtain.\\
This approach is referred to as semi-supervised, since on one hand, no labeled data is required to obtain the PC scores from acceleration measurements, but on the other hand, some labeled examples are still required to physically interpret the PC scores and to link them to specific maneuvers. This is achieved using the training examples from the labeled dataset. \autoref{fig:maneuver_separation} shows that principal components 1 and 2 contain a lot of information on the riding underground and speed. In the underground separation plot (a), all examples of the uneven underground are located in the left branch and the examples of even undergrounds are clustered on the right. As a result, it is very easy for a classification algorithm to correctly classify the underground from the data, for example by placing a decision boundary in between both categories. From the riding speed separation plot (b), it can be concluded that an increasing riding speed is reflected by a decrease in PC 1. The speed could therefore be classified into bins of measured speeds depending on a corresponding principal component range. While information about the rider of each labeled measurement and their respective weight is available in the dataset, no combination of two principal components was found which clearly separates these classes visually.\\

\subsection{Discrete maneuver classification example} \label{sec:discrete_maneuver_identification}
\begin{figure*}
	\begin{center}
		\subfloat[Scattering based underground classification]{
			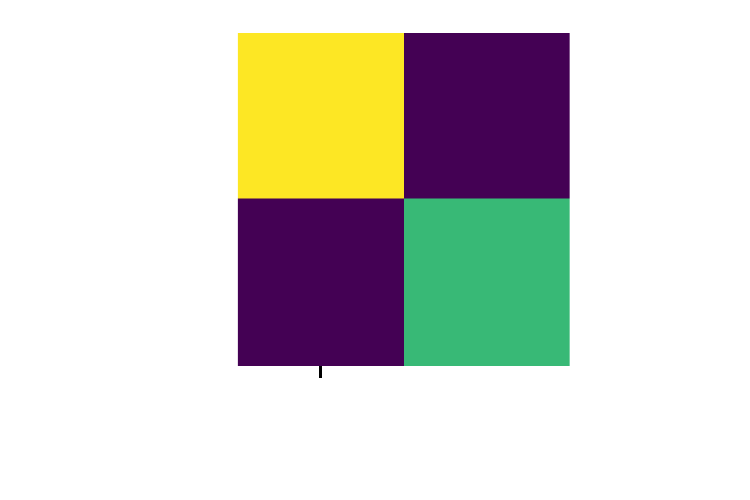
		}
		\subfloat[Fast fourier based underground classification]{
			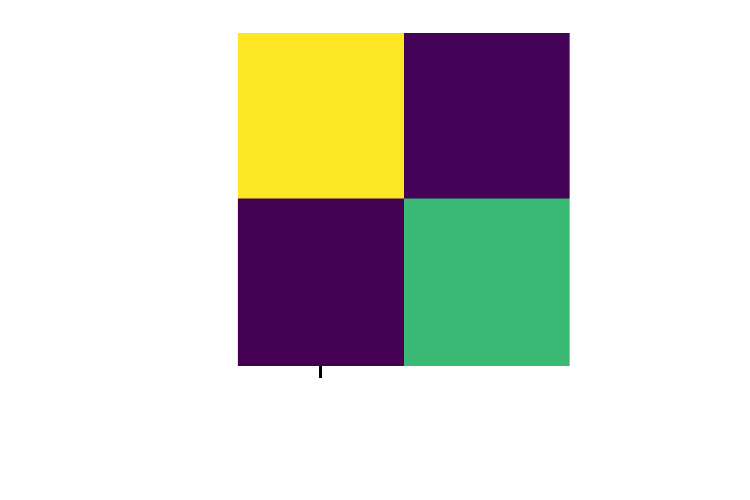
		}
		\caption[Underground classification]{Using the scattering transform based representation, the underground is classified with a perfect accuracy of \SI{100}{\%}, while the fast Fourier transform leads to an almost perfect accuracy of \SI{99.5}{\%}.}
		\label{fig:undeground_classification}
	\end{center}
\end{figure*}

\begin{figure*}
	\begin{center}
		\subfloat[Scattering based speed classification]{
			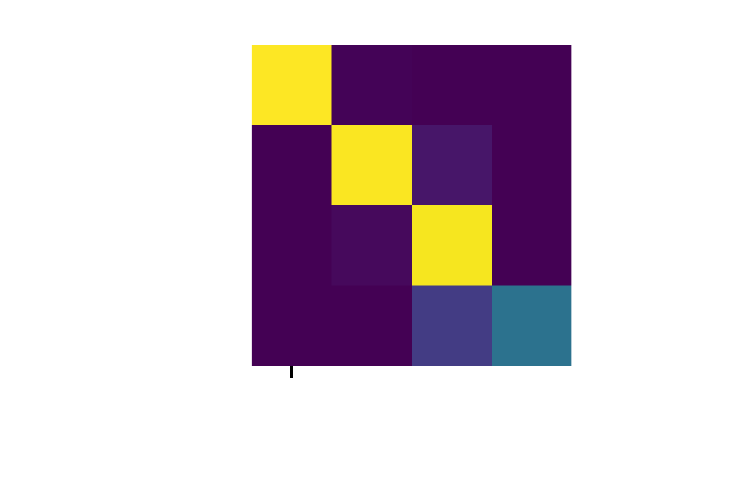
		}
		\subfloat[Fast Fourier based speed classification]{
			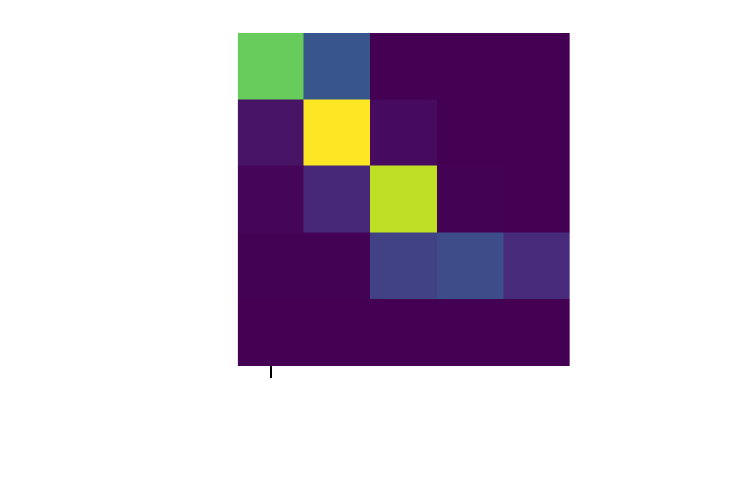
		}
		\caption[Speed classification]{Using the scattering transform, the riding speed can be classified with an accuracy of \SI{92.6}{\%}. Here, no samples are incorrectly classified by more than one class and the erroneous samples are mostly samples at \SI{20}{\kilo\meter/\hour} that are falsely classified as \SI{15}{\kilo\meter/\hour}. Using the fast Fourier based representation, falsely classified samples exist between all classes, leading to an accuracy of only \SI{78.3}{\%}. Since samples of \SI{25}{\kilo\meter/\hour} exist in the even underground training data, but not in the testing data, some samples with a speed of \SI{20}{\kilo\meter/\hour} are incorrectly attributed to this range for the Fast Fourier based representation, but not for the Scattering based variant.}
		\label{fig:speed_classification}
	\end{center}
\end{figure*}

\begin{figure*}
	\begin{center}
		\subfloat[Rider classification scattering]{
			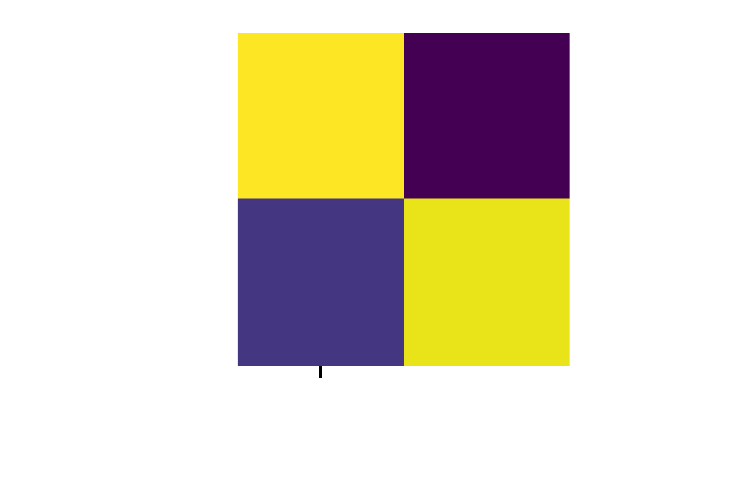
		}
		\subfloat[Rider classification fast fourier]{
			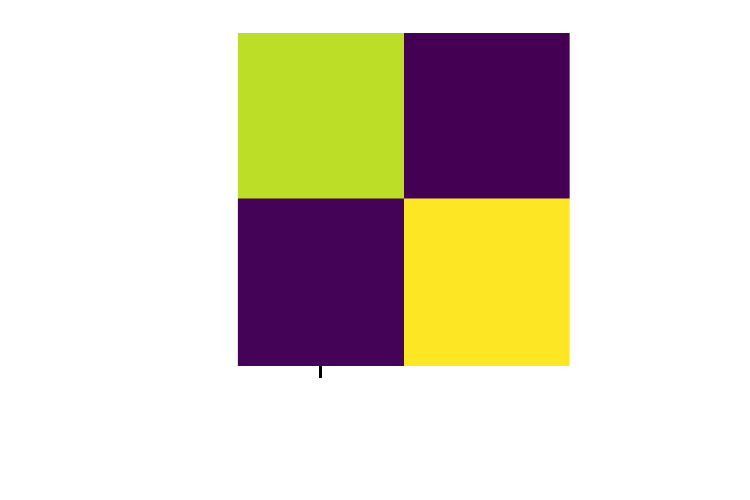
		}
		\caption[Rider classification]{The rider classification based on the scattering representation achieves an accuracy of \SI{90.9}{\%}, with most incorrect predictions stemming from measurement data of rider 2. The fast Fourier based representation yields a near perfect accuracy of \SI{98.7}{\%}, where false predictions are evenly distributed between both riders.}
		\label{fig:rider_classification}
	\end{center}
\end{figure*}

With the knowledge that underground and speed information are available in the principal component space, a classification model can be created, which predicts labels from PC scores. These labels can be categorical, e.g. even or cobblestone underground, discrete numerical, e.g. \SI{10}{\kilo\meter/\hour} or \SI{15}{\kilo\meter/\hour}, or a combination of both, e.g. rider 1 with weight \SI{66}{\kilogram} or rider 2 with weight \SI{83}{\kilogram}. For this model, a k-Nearest-Neighbors classifier \cite{IntroStats} is chosen, where the label of a new example is predicted based on the most common labels among its $k$ nearest neighboring points, selected using the euclidean norm over all 10 principal component axes. The parameter $k=20$ was selected arbitrarily based on the total number of points in the training dataset. In more complex classification tasks, it can be estimated using cross validation \cite{IntroStats}. Once the classification model is parameterized, it is employed to predict all examples from the labeled testing dataset. Again, a comparison is made between the reduced acceleration representation based on the scattering transform and the Fast Fourier transform.\\
The underground classification results are depicted as a confusion matrix in \autoref{fig:undeground_classification}. It shows that the scattering based classifier is able to perfectly predict all undergrounds, while the Fast Fourier based classifier reaches an almost perfect accuracy of \SI{99.5}{\%}. \autoref{fig:speed_classification} shows the classification of riding speed categories. Here, the scattering based approach yields an accuracy of \SI{92.6}{\%}, while the Fast Fourier based model only reaches \SI{78.3}{\%}. It is also worth noting that for even underground data, training data of up to \SI{25}{\kilo\meter/\hour} is available, while the testing dataset only includes examples up to \SI{20}{\kilo\meter/\hour}, resulting from a different measuring location. While the scattering based approach never falsely classifies a testing sample as \SI{25}{\kilo\meter/\hour}, this problem occurs for the Fast Fourier based classifier. For the last classification case, \autoref{fig:rider_classification} shows the rider classification results. Here, the Fast Fourier based model outperforms the Scattering transform with a near perfect accuracy of \SI{98.7}{\%} compared to \SI{90.9}{\%}. 

\section{Discussion} \label{sec:discussion}
This paper demonstrates the feasibility of the presented fatigue monitoring approach using a demonstrator. For the transfer to real vehicle fleets, further challenges need to be addressed.\\
Special care has to be taken regarding the precise positioning and calibration of sensors. Fortunately, quality control measures can be employed to compare the sensor data between different standard vehicles for common test maneuvers. Furthermore, manufacturing processes lead to small differences in material properties and geometry between individual vehicles. The statistics of these differences can be incorporated in multiple ways. When local \textsc{Wöhler}-curves are available for the monitored strain positions, these can directly replace the fictitious \textsc{Wöhler}-curves used in this paper in order to relate the monitoring damage sums to the corresponding statistics of failure. Here, it might be useful to incorporate an additional correction factor into the regressor in order to ensure that predicted damage sums are unbiased. The corresponding correction factor can be identified using cross validation on the training dataset. In cases where a local \textsc{Wöhler}-curve is not available, fictitious damage sums can be interpreted as fatigue indicators, which steadily increase with vehicle usage. If the corresponding fleet is sufficiently large, these indicator values can be correlated with actual component failures which are observed during maintenance, leading to a monitoring strategy that automatically improves in accuracy over time. For each component, the probability of failure can be modeled as a normal distribution with a corresponding mean and standard deviation formulated in terms of the fatigue indicators.\\
In \autoref{sec:discrete_maneuver_identification}, it was shown that labeled data can be used to identify a given set of predefined maneuvers. This was mainly done to show that maneuver information is accessible through the reduced acceleration data representation, but might be unsatisfactory when the expected number of relevant maneuvers is very high. In this case, it might be feasible to use clustering techniques like k-means clustering or gaussian mixture models to automatically identify clusters from unlabeled data. These methods might be especially promising when large differences in the overall system behavior are expected between different maneuvers.\\
The presented approach is fundamentally designed to be parameterized using only measurement data and therefore does not require a finite element solution to function. Still, information from an FE model can contribute to the accuracy of this approach by identifying fatigue critical strain sensor locations from simple load cases, or through analyzing mode shapes in order to find anti-nodes to position acceleration sensors. Unfortunately, no FE model of the eBike was available at the time of sensor placement in this study, therefore further improvements can be expected by taking such information into account. The extrapolation of monitoring data from a finite number of discrete strain sensors to full-field strain information of the eBike frame is therefore also beyond the scope of this study.\\
Finally, the relation between the presented monitoring strategy and structural health monitoring shall be discussed. Structural health monitoring fundamentally assumes that system failure is preceded by a change in the dynamic system characteristics. Therefore, the aim of monitoring in SHM is to identify this change, so that appropriate measures can be taken. In contrast, the strategy presented in this paper assumes that fatigue damage builds up over a long period of time without influencing the system behavior, before leading to abrupt failure. As a result, this failure can only be predicted by accumulating fatigue damage sums over the entire history of vehicle usage. At the same time, system monitoring through the presented approach is no longer valid once a significant change in the system behavior has occurred, since identical system behavior is the fundamental assumption of the standard and reference vehicle setup. Which approach yields better predictions, between SHM and this paper, depends largely on the properties of the system or its components being monitored. However, both approaches rely on similar sensor setups, so that they can also be used in conjunction, i.e. scheduling maintenance after either a critical fatigue damage sum is reached or a change in the system behavior is detected.

\section{Conclusions} \label{sec:conclusions}
The virtual sensing approach presented in this paper enables fatigue monitoring for vehicle fleets with a reduced sensor setup, using a combination of feature extraction and dimensionality reduction. This is enabled by a reference approach, where all vehicles in a fleet have a given set of standard acceleration sensors, which are used to identify the dynamic state of the vehicle. These standard measurements are extended by additional reference sensors, which only have to be installed at one reference vehicle and provide local strain responses, necessary for the fatigue assessment. A sensor equipped eBike served as the demonstrator for this approach, which was used to obtain an unlabeled and a labeled dataset for model parameterization and validation. A low dimensional representation of the acceleration measurements constitutes the core of both application cases. In the first study, it was shown that a strong correlation between the reduced acceleration data representation and the fatigue damage sums of concurrent strain measurements exist, which was subsequently used in a fatigue damage regression. The second application used a semi-supervised approach to physically interpret the principal component scores in the reduced representation, enabling a maneuver classification. Two methods for feature extraction, namely the scattering transform and the Fast Fourier transform, where compared in multiple applications, while principal component analysis was used for the dimensionality reduction task. The scattering transform outperformed the FFT based model in most regression examples, yielding more accurate fatigue damage predictions per time sequence in unlabeled and labeled datasets. In the maneuver identification tasks, both models classified the underground very well and the scattering transform outperformed the Fast Fourier based model in terms of driving speed estimation, while falling short in the rider category. In the discussion, different challenges related to the transfer of this monitoring strategy to real vehicle fleets were highlighted and strategies to overcome these challenges were explored. The implementation of these strategies is subject of further investigation.

\section{Declaration of Competing Interrest}
The authors declare that they have no known competing financial interests or personal relationships that could have appeared to influence the work reported in this paper.

\section{Acknowledgments}
\begin{table}[H]
 \begin{tabular}{m{8cm} m{1.5cm} m{5cm}  }
  \includegraphics[height=20mm]{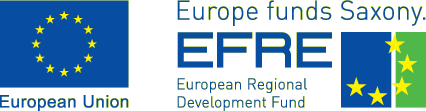} &   \includegraphics[height=20mm]{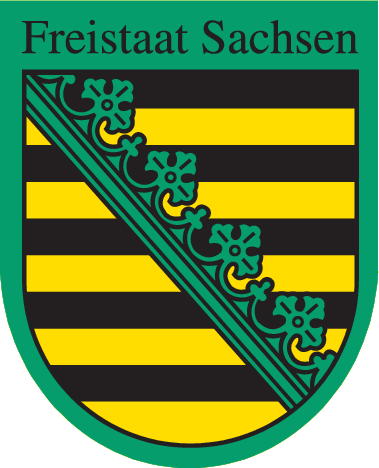} & 
This measure is co-financed with tax revenues on the basis of the budget adopted by the members of the Saxon State Parliament.
 \end{tabular}
\end{table}	
The authors gratefully acknowledge the GWK support for funding this project by providing computing time through the Center for Information Services and HPC (ZIH) at TU Dresden.

\bibliography{Bibliography}


\end{document}